\documentclass[a4paper,fleqn,usenatbib]{mnras}

\usepackage{newtxtext,newtxmath}
\usepackage[T1]{fontenc}
\usepackage{ae,aecompl}

\usepackage{graphicx}	% Including figure files
\usepackage{amsmath}	% Advanced maths commands
\usepackage{amssymb}	% Extra maths symbols
\graphicspath{{./plots/}}

\newcommand{\msun}{M$_{\odot}$}

\newcommand{\kms}{km~s$^{-1}$}
\newcommand{\ergs}{erg s$^{-1}$}
\newcommand{\Ha}{H$\alpha$}

\newcommand{\HII}{H~{\sc ii}}
\newcommand{\HeI}{He~{\sc i}}
\newcommand{\OI}{O~{\sc i}}
\newcommand{\Oneb}{[O~{\sc i}]}

\newcommand{\NaI}{Na~{\sc i}}

\newcommand{\SiII}{Si~{\sc ii}}

\newcommand{\CaII}{Ca~{\sc ii}}

\newcommand{\FeII}{Fe~{\sc ii}}

\newcommand{\Cofs}{$^{56}$Co}
\newcommand{\Nifs}{$^{56}$Ni}
\newcommand{\mej}{$M_\mathrm{ej}$}
\newcommand{\ek}{$E_\mathrm{k}$}
\newcommand{\vph}{$v_\mathrm{ph}$}
\newcommand{\vsc}{$v_\mathrm{sc}$}
\newcommand{\lp}{$L_\mathrm{p}$}
\newcommand{\trise}{$t_{-1/2}$}
\newcommand{\tdecay}{$t_{+1/2}$}

\newcommand{\lam}{$\lambda$}
\newcommand{\loglp}{$\mathrm{log}\left(L_\mathrm{p}\right)$}
\newcommand{\tmax}{$t_\mathrm{max}$}
\newcommand{\Eh}{$E\left(B-V\right)_\mathrm{host}$}
\newcommand{\Emw}{$E\left(B-V\right)_\mathrm{MW}$}
\newcommand{\Etot}{$E\left(B-V\right)_\mathrm{tot}$}
\newcommand{\E}{$E\left(B-V\right)$}
\newcommand{\mni}{$M_\mathrm{Ni}$}

\newcommand{\tp}{$t_\mathrm{p}$}

\newcommand{\mzams}{$M_\mathrm{ZAMS}$}

\newcommand{\taum}{$\tau_\mathrm{m}$}
\newcommand{\dlate}{$\delta m_\mathrm{100}$}
\newcommand{\texp}{$t_\mathrm{exp}$}
\newcommand{\md}{mag d$^{-1}$}
\newcommand{\mhe}{$M_\mathrm{He}$}

\title[ SE-SNe distributions and properties]{Investigating the properties of stripped-envelope supernovae; what are the implications for their progenitors?}

\author[S. J. Prentice]{S. J. Prentice,$^{1,2}$\thanks{E-mail: sipren.astro@gmail.com}, C. Ashall$^{2,3}$, P. A. James$^{2}$, L. Short$^{2}$, P. A. Mazzali$^{2,4}$, D. Bersier$^{2}$, 
\newauthor P. A. Crowther$^{5}$, C. Barbarino$^{6}$, T.-W. Chen$^{7}$, C. M. Copperwheat$^{2}$, M. J. Darnley$^{2}$, 
\newauthor L. Denneau$^{8}$, N. Elias-Rosa$^{9,10}$, M. Fraser$^{11}$, L. Galbany$^{12}$, A. Gal-Yam$^{13}$,   
\newauthor J. Harmanen$^{14}$, D. A. Howell$^{15,16}$, G. Hosseinzadeh$^{15,16,24}$, C. Inserra$^{17}$,  E. Kankare$^{1}$, 
\newauthor  E. Karamehmetoglu$^{6}$, G. P. Lamb$^{2,18}$, M. Limongi$^{19,20}$, K. Maguire$^{1}$, C. McCully$^{15,16}$
\newauthor F. Olivares E.$^{21,22}$, A. S. Piascik$^{2}$, G. Pignata$^{23,21}$, D. E. Reichart$^{25}$, A. Rest$^{26,27}$      
\newauthor T. Reynolds$^{14}$, \'O. Rodr\'iguez$^{23,21}$, J. L. O. Saario$^{28,29,30}$, S. Schulze$^{13}$, S. J. Smartt$^{1}$,        
\newauthor  K. W. Smith$^{1}$, J. Sollerman$^{6}$, B. Stalder$^{31}$, M. Sullivan$^{17}$, F. Taddia$^{6}$, S. Valenti$^{32}$,  
\newauthor S. D. Vergani$^{33}$, S. C. Williams$^{34}$, D. R. Young$^{1}$
\\
$^{1}$Astrophysics Research Centre, School of Mathematics and Physics, Queen's University Belfast, BT7 1NN, UK\\
$^{2}$Astrophysics Research Institute, Liverpool John Moores University, IC2, Liverpool Science Park, 146 Brownlow Hill, \\  Liverpool L3 5RF, UK\\
$^{3}$Department of Physics, Florida State University, Tallahassee, FL 32306, USA\\
$^{4}$Max-Planck-Institut f{\"u}r Astrophysik, Karl-Schwarzschild-Str. 1, D-85748 Garching, Germany\\
$^{5}$Department of Physics \&\ Astronomy, University of Sheffield, Sheffield, UK \\
$^{6}$The Oskar Klein Centre, Department of Astronomy, Stockholm Univsersity, AlbaNova, SE-106 91 Stockholm , Sweden\\
$^{7}$Max-Planck-Institut f{\"u}r Extraterrestrische Physik, Giessenbachstra\ss e, 85748, Garching, Germany\\
$^{8}$Institute for Astronomy, University of Hawai'i, 2680 Woodlawn Drive, Honolulu, HI 96822, USA\\
$^{9}$Institute of Space Sciences (ICE, CSIC), Campus UAB, Carrer de Can Magrans s/n, 08193 Barcelona, Spain\\
$^{10}$Institut d'Estudis Espacials de Catalunya (IEEC), C/ Gran Capit\'a 2-4, Edif. Nexus 201, 08034 Barcelona, Spain\\
$^{11}$School of Physics, O'Brien Centre for Science North, University College Dublin, Belfield, Dublin 4, Ireland\\
$^{12}$PITT PACC, Department of Physics and Astronomy, University of Pittsburgh, Pittsburgh, PA 15260, USA\\
$^{13}$Department of Particle Physics and Astrophysics, Weizmann Institute of Science, Rehovot 76100, Israel\\
$^{14}$Tuorla Observatory, Department of Physics and Astronomy, FI-20014 University of Turku, Finland\\
$^{15}$Las Cumbres Observatory, 6740 Cortona Dr. Suite 102, Goleta, CA, USA 93117\\
$^{16}$University of California, Santa Barbara, Department of Physics, Broida Hall, Santa Barbara, CA, USA 93111\\
$^{17}$Department of Physics and Astronomy, University of Southampton, Southampton, SO17 1BJ, UK\\
$^{18}$Department of Physics and Astronomy, University of Leicester, University Road, Leicester LE1 7RH, UK\\
$^{19}$Istituto Nazionale di Astrofisica-Osservatorio Astronomico di Roma, Via Frascati 33, I-00040, Monteporzio Catone, Italy\\
$^{20}$Kavli Institute for the Physics and Mathematics of the Universe, Todai Institutes for Advanced Study, the University of Tokyo,\\ Kashiwa, 277-8583, Japan\\
$^{21}$Millennium Institute of Astrophysics, Nuncio Monse\~nor S\'otero Sanz 100, Providencia, Santiago, Chile\\
$^{22}$Departamento de Astronom\'{\i}a, Universidad de Chile, Camino el Observatorio 1515, Santiago, Chile\\
$^{23}$Departamento de Ciencias Fisicas, Universidad Andres Bello, Avda. Republica 252, Sazi\'e, 2320, Santiago, Chile\\
$^{24}$Harvard-Smithsonian Center for Astrophysics, 60 Garden Street, Cambridge, MA 02138-1516, USA\\
$^{25}$University of North Carolina 269 Phillips Hall, CB 3255 Chapel Hill, NC 27599\\
$^{26}$Space Telescope Science Institute, 3700 San Martin Drive, Baltimore, MD 21218, USA\\
$^{27}$Department of Physics and Astronomy, Johns Hopkins University, Baltimore, MD 21218, USA\\
$^{28}$Nordic Optical Telescope, Apartado 474, E-38700 Santa Cruz de La Palma, Spain\\
$^{29}$Koninklijke Sterrenwacht van Belgi\"e, Ringlaan 3, 1180 Brussels, Belgium\\
$^{30}$Institute of Astronomy, KU Leuven, Celestijnenlaan 200D, 3001 Leuven, Belgium\\
$^{31}$LSST, 950 N Cherry Ave, Tucson, AZ 95719\\
$^{32}$Department of Physics, University of California, Davis, CA 95 616, USA\\
$^{33}$GEPI, Observatoire de Paris, PSL Universite, CNRS, 5 Place Jules Janssen, F-92190 Meudon, France\\
$^{34}$Physics Department, Lancaster University, Lancaster, LA1 4YB, UK\\
}

\date{Accepted XXX. Received YYY; in original form ZZZ}

\pubyear{2018}

\begin{document}
\label{firstpage}
\pagerange{\pageref{firstpage}--\pageref{lastpage}}
\maketitle

\begin{abstract}
We present observations and analysis of 18 stripped-envelope supernovae observed during 2013 -- 2018.
This sample consists of 5 H/He-rich SNe, 6 H-poor/He-rich SNe, 3 narrow lined SNe Ic and 4 broad lined SNe Ic.
The peak luminosity and characteristic time-scales of the bolometric light curves are calculated, and the light curves modelled to derive \Nifs\ and ejecta masses (\mni\ and \mej).
Additionally, the temperature evolution and spectral line velocity-curves of each SN are examined. 
Analysis of the \Oneb\ line in the nebular phase of eight SNe suggests their progenitors had initial masses $<20$ \msun.
The bolometric light curve properties are examined in combination with those of other SE events from the literature. The resulting dataset gives the \mej\ distribution for 80 SE-SNe, the largest such sample in the literature to date, and shows that SNe Ib have the lowest median \mej, followed by narrow lined SNe Ic, H/He-rich SNe, broad lined SNe Ic, and finally gamma-ray burst SNe. SNe Ic-6/7 show the largest spread of \mej\, ranging from $\sim 1.2 - 11$ \msun, considerably greater than any other subtype. 
For all SE-SNe $<$\mej$>=2.8\pm{1.5}$ \msun\ which further strengthens the evidence that SE-SNe arise from low mass progenitors which are typically $<5$ \msun\ at the time of explosion, again suggesting \mzams\ $<25$ \msun.   
The low $<$\mej$>$ and lack of clear bimodality in the distribution implies $<30$ \msun\ progenitors and that envelope stripping via binary interaction is the dominant evolutionary pathway of these SNe.%240 words

\end{abstract}

\begin{keywords}
Supernovae:general
\end{keywords}

%%%%%%%%%%%%%%%%%%%%%%%%%%%%%%%%%%%%%%%%%%%%%%%%%%

%%%%%%%%%%%%%%%%% BODY OF PAPER %%%%%%%%%%%%%%%%%%

\section{Introduction}\label{sec:intro}
Stripped-envelope supernovae (SE-SNe, SNe Ib/c) are a subset of core-collapse events that arise from stars almost completely deficient in helium and/or hydrogen at the time of explosion.
These events are named on the basis of their spectroscopic appearance. The presence of strong \Ha\ and other Balmer lines leads to a SN IIb event, those without clear H signatures but are strong in He are SNe Ib, and those without traces of He are SNe Ic \citep[see][]{Filippenko1995,Matheson2001}. Although there are potentially SNe Ic with traces of He in the ejecta \citep[SN 2016coi;][]{Yamanaka2017,Prentice2018}, the vast majority are incompatible with He \citep{Modjaz2016,Taddia2018}.

In the classification scheme of \cite{Prentice2017} the sequence of He-rich SNe was refined into four subclasses, each with progressively less evidence for H in the ejecta; SNe IIb, SNe IIb(I), SNe Ib(II), and SNe Ib.
For SNe Ic the classification was based upon quantifying the degree of line blending. Based upon a set of common lines found in the optical spectra the SNe were classified from narrow lined to broad-lined as; SNe Ic-7, SNe Ic-6, SNe Ic-5, SNe Ic-4, and SNe Ic-3. SNe Ic-5 are the transition between broad lined (Ic-3/4) and narrow lined (Ic-6/7) events. This classification scheme is adopted through this work and can be approximated with the traditional taxonomy by assuming
\begin{itemize}
	\item{IIb + IIb(I) -- H/He-rich SNe $\sim$ IIb} 
	\item{Ib + Ib(II) -- H-poor/He-rich $\sim$ Ib}
	\item{Ic-5/6/7 -- narrow-lined SNe $\sim$ Ic}
    \item{Ic-3/4 -- non-GRB Ic-BL}
\end{itemize}

The quantity of data on these SE types has rapidly increased in the last few years thanks to large data releases \citep[e.g.,][]{Bianco2014,Taddia2015,Stritzinger2018b}, 
which allows more in-depth study of their light curves \citep{Drout2011,Lyman2016,Taddia2018, Prentice2016} and spectroscopic properties \citep{Liu2016,Modjaz2016,Prentice2017,Fremling2018}.

SE SNe typically rise on timescales between 10 and 20 days, though some extreme events can be nearly 40 days \citep[e.g.,\  SN 2011bm, iPTF15dtg;][]{Valenti2012,Taddia2016}.
The most extremely energetic events are defined by broad absorption lines in their spectra \citep{Iwamoto1998}, with some associated with X-ray flashes (XRF) and gamma-ray bursts (GRBs). GRB-SNe can have large ejecta masses, \mej\ $\sim 8 - 12$ \msun, and kinetic energy \ek\ that can exceed $10^{52}$ erg \citep[see][]{Nakamura2001,Mazzali2003,Mazzali2006b,Olivares2015}.
The extreme \ek\ of these events is thought to come from the rotational energy of a compact object \citep{Woosley1993b,Mazzali2014,Ashall2017}.
There is also evidence that SE events are inherently aspherical \citep[e.g.][]{Mazzali2005,Maeda2008,Wang2008,Tanaka2009,Chornock2011,Mauerhan2015,Stevance2016,Stevance2017}.

Progenitors of SE-SNe have been detected in pre-explosion imaging or colour excess in post-explosion in photometry after the SN has faded, mostly for H-rich SE-SNe; SN 1993J \citep{Aldering1994,Smartt2009}, SN 2008ax \citep{Crockett2008,Arcavi2011,Folatelli2015}, SN 2011dh \citep{Maund2011}, SN 2013df \citep{VanDyk2014}, and SN 2016gkg \citep{Kilpatrick2017,Tartaglia2017}. In all cases the progenitor star was estimated to have \mzams\ $< 20$ \msun.
The progenitor of SN Ib iPTF13bvn was estimated to have \mzams\ $=10 - 20$ \msun, and a possible binary companion \citep[see][]{Cao2013,Eldridge2016,Fremling2014,Bersten2014}. 

There is currently only one progenitor candidate for SNe~Ic, a \mzams\ $=47-80$ \msun\ star for Ic-7 SN 2017ein, however the identification of the actual progenitor is uncertain and quite possibly the object identified is a compact star cluster \citep{VanDyk2018}. 
The progenitors of SNe Ic are plausibly WO and WC Wolf-Rayet stars at the time of explosion \citep{Crowther2007}. They are UV bright but relatively dim in the optical and may be further attenuated by local dust formation \citep{Crowther2003}. This leads to difficulty in locating them in pre-explosion imaging \citep{Smartt2009,Eldridge2013}.

In the absence of a progenitor candidate, another method is to estimate the age of the stellar population at the site of the explosion. \cite{Anderson2012} found that SNe Ib/c are more often associated with \Ha\ emission and interpreted this as evidence of stars from young populations and are therefore more massive than the typical progenitor for SNe II. \cite{Galbany2016} examined the host of nearby SNe and suggested that SNe Ib are the result of binary stripping but that some SNe~Ic are likely single massive stars.
Similar results were found by \cite{Kuncarayakti2018}, some SE-SNe favour low mass progenitors but their stellar populations are typically younger. The stellar populations at the location of SNe Ic are often calculated to be the youngest, hence their progenitors are assumed to be the most massive \citep[e.g.,][]{Galbany2018}. \cite{Maund2018} estimates that these SNe arise from stars $> 30$ \msun.
In the cases of specific SNe, \cite{Maund2016} estimates that the progenitor of SNe Ic-7 2007gr \citep[\mej\ <2 \msun ;][]{Mazzali2009} to be \mzams\ $\sim 30$ \msun.
However, \cite{Crowther2013} cautions against inferring stellar age from \HII\ regions, as these can survive for tens of millions of years and give rise to several generations of massive stars and limits mass estimates in these regions to $>12$ \msun.
Regardless, there is a propensity for SNe Ib and Ic to be found nearer to \HII\ regions compared with SNe II.

For a single star, mass loss is an increasing function of initial mass, metallicity, and rotation. 
SE-SNe are $\sim 30$ percent of all core-collapse events in the local Universe and occur at relative volume-limited rate of $N_\mathrm{IIb}\approx N_\mathrm{Ib} \approx 0.8N_\mathrm{Ic}$ \citep{Shivvers2017}.
These rates are not compatible with a purely mass-dominated regime.
However, mass loss can also occur through envelope stripping via interaction with a close binary companion, allowing stars with lower initial masses than in the single star scenario to be stripped \citep[e.g.,][]{Nomoto1995}. 
The binary fraction of massive stars close enough to undergo mass transfer is 70\%\ in the Milky Way \citep{Sana2012}, so this is a physical process which affects most massive stars. There have been direct detections of binary companions to three stripped-supernovae; SNe 1993J, 2011dh, and 2001ig \citep{Maund2004,Folatelli2014,Fox2014,Maund2015,Ryder2018}. Upper limits have been placed on the mass of companions for Ic-6 SN 1994I \citep[10 \msun;][]{VanDyk2016} and Ic-4 SN 2002ap \citep[8--10 \msun, also used to limit the progenitor of SN 2002ap to \mzams\ $<23$ \msun;][]{Zapartas2017}.

This work presents photometric and spectroscopic observations of 18 SE-SNe, observed between 2013 -- 2018 as part of the Public ESO Spectroscopic Survey of Transient Objects (PESSTO), extended PESSTO (ePESSTO) \citep{Smartt2015}, and a Liverpool Telescope SE-SN follow up campaign, and the Las Cumbres Observatory Supernova Key Project.
The observations and data reduction methods are presented in Section~\ref{sec:obs}. Section~\ref{sec:lcs} presents the multi-colour light curves, the bolometric light curve and modelling, and the temperature evolution of the SNe. In Section~\ref{sec:spectra} the maximum light and nebular phase spectra are shown along with analysis of the line velocities. The results are set into context against other SE-SNe in Section~\ref{sec:comp}, and in Section~\ref{sec:masses} \Nifs\ and ejecta mass distributions are investigated. The conclusions are summarised in Section~\ref{sec:conclusion}.

\section{Data collection, reduction, and calibration}\label{sec:obs}
\begin{table*}
	\centering
	\caption{General information on the SNe, host galaxies, and reddening }
	\begin{tabular}{lcccccccc}
    \hline
	SN & Type &$\alpha$	& $\delta$	& Gal type$^{a}$	& $z$ & $\mu$ & $E\left(B-V\right)_\mathrm{MW}$ & $E\left(B-V\right)_\mathrm{host}$ \\
	& & (J2000)&  & & & [mag] &[mag]&[mag] \\
    \hline
    
    2013F	& Ic-6  &25:48:25.03 & -41:19:56.21 &  SABbc & 0.005 & 31.29 & 	0.018 &  1.4$\pm{0.2}$  \\
    
	2013bb	& IIb(I) & 14:12:13.96 & +15:50:31.49 & SAB(s)bc  & 0.018 & 34.52 & 	0.014 &  0.3$\pm0.1$\\
    
    2013ek	& Ib  &20:57:53.90 & -51:52:24.49 & SBc  & 0.016 & 33.50 & 0.033 &  0.04$\pm0.01$  \\

	2015ah &  Ib  &23:00:24.63 & +01:37:36.80 & SABcd & 0.016 & 33.86 &0.071 &$0.02\pm{0.01}$ \\
    
    2015ap &  Ib  &02:05:13.32 & +06:06:08.39 & SBd & 0.011 & 33.27 & 0.037 & negligible \\
    
	2016P & Ic-6*  &13:57:31.10 &  +06:05:51.00 & SBbc & 0.015 & 34.17 & 0.024 & $0.05\pm{0.02}$\\
    
	2016frp &  Ib  &00:21:32.54 & -05:57:24.29 & - & 0.027 & 35.23 &0.031 & negligible \\
    
    2016gkg &  IIb  &01:34:14.46 & -29:26:25.00 & SB(rs)bc & 0.005 & 31.22  &0.017 & 0.05$\pm{0.02}$  \\
    
	2016iae &  Ic-7  &04:12:05.53 & -32:51:44.75 & SBb pec  & 0.004 & 31.19 & 0.014 & $0.65\pm{0.2}$\\
    
    2016jdw$\dag$	& Ib  & 13:16:19.62 & +30:40:32.67 & -  & 0.019 & 34.66 & 	0.01 &  negligible  \\
 
	2017bgu$\dag$	& Ib  & 16:55:59.47 & +42:33:36.01 & -  & 0.009 & 33.20 & 	0.029 &  $0.02\pm{0.01}$  \\
    
    2017dcc &  Ic-3  &12:49:04.89 & -12:12:22.42 & - & 0.025 & 35.06 & 0.041 & negligible\\
    
    2017gpn$\dag$	& IIb(I)  & 03:37:45.26 & +72:31:58.70 & SAB(s)b: pec  & 0.007 & 32.60 & 	0.3 &  negligible  \\
    
    2017hyh &  IIb(I)  &07:10:41.07 & +06:27:41.40 & - & 0.012 & 33.54 & 0.118 & $0.04\pm{0.01}$\\
    
    2017ifh$\dag$	& Ic-4  & 06:35:03.56 & +50:26:27.90 & -  & 0.039 & 35.97 & 	0.117 &  $0.05\pm{0.02}$  \\
 
	2017ixz	& IIb  & 07:47:03.03 & +26:46:25.77 & -  & 0.024 & 34.98 & 0.033 &  negligible  \\
 
	2018ie$\dag$	& Ic-4  & 10:54:01.06 & -16:01:21.40 & SB(rs)c  & 0.014 & 33.35 & 	0.06 &  $0.02\pm{0.01}$  \\
 
	2018cbz$\dag$	& Ic-4*  & 13:41:18.61 & -04:20:46.54 & SBd   & 0.022 & 35.03 & 0.03 &  $0.03\pm{0.025}$  \\
    
    \hline
    \multicolumn{8}{l}{$^a$ See Appendix~\ref{sec:hosts} for discussion on some of the host environments}\\
    \multicolumn{8}{l}{*Ambiguous classification}\\
    \multicolumn{8}{l}{$\dag$Observed independently of PESSTO/ePESSTO}\\
    
	\end{tabular}
	\label{tab:SNe}
\end{table*}

The SNe, listed in Table~\ref{tab:SNe}, were discovered by a variety of different sources over five years.
Subsequent observations were obtained from various locations and instruments, as follows:

\begin{itemize}
	\item{The 3.6~m ESO New Technology Telescope (NTT) at La Silla, Chile, using the ESO Faint Object Spectrograph and Camera (v.2) (EFOSC2) for both spectroscopy and $BVRI$ imaging.}
    \item{The 2.0 m Liverpool Telescope \citep[LT;][]{Steele2004} at the Roque de los Muchachos Observatory, La Palma, Spain. $BVugriz$ photometric data was obtained using IO:O, while spectroscopic data were collected using the Spectrograph for the Rapid Acquisition of Transients \citep[SPRAT;][]{Piascik2014}.}
    \item{The Las Cumbres Observatory \citep[LCO;][]{Brown2013} network of robotic telescopes. Spectroscopy was obtained using the Floyds spectrographs and $BVg_p r_p i_p$ imaging from the Spectral cameras on the LCO 2.0 m telescopes at Hawaii and the Siding Spring Observatory. Photometry was also obtained from the 1 m telescopes at the McDonald Observatory, the Cerro Tololo Inter-American Observatory (CTIO) and the South African Astronomical Observatory (SAAO).}
	\item{Near-infrared $JHK$ imaging from NOTcam on the Nordic Optical Telescope (NOT), operated by the Nordic Optical Telescope Scientific Association at the Observatorio del Roque de los Muchachos. }
\end{itemize}
\begin{itemize}	
    \item{The Gamma-Ray Burst Optical/Near-Infrared Detector \citep[GROND;][]{Greiner2008}, a 7-channel imager that collects multi-colour photometry simultaneously with $g'r'i'z'JHKs$ bands, mounted at the 2.2 m MPG telescope at ESO La Silla Observatory in Chile.}
    \item{The twin ATLAS 0.5 m telescope system based on the Haleakala and Mauna Loa, Hawaii, USA \citep{Tonry2018}. The ATLAS $c$ and $o$ filters are approximately $g+r$ and $r+i$ respectively.}
    \item{The Wide-field Spectrograph (WiFeS) on the Australian National University 2.3 m telescope at the Siding Spring Observatory (on behalf of PESSTO).}
    \item{Images were acquired with the $g'$ filter mounted on the  60cm PROMPT3 telescope (pixel size of 1.4") and with the Luminance filter mounted on the 80cm PROMPT7 telescope (pixel size of 0.66"). Both telescopes are located at Cerro Tololo Inter-American (CTIO) observatory, Chile. Observations by the Chilean Automatic Supernova Search \citep[CHASE;][]{Pignata2009}.}
    \item{$BVRI$ photometry using the ANDICAM-CCD on the SMARTS 1.3~m CTIO observatory, Chile.}
    \item{ The Deep Imaging Multi-Object Spectrograph \citep[DEIMOS;][]{Faber2003}
on the W. M. Keck Observatory, Haleakala, as part of LCO Supernova Key Project}
\end{itemize}

Alignment, stacking, aperture photometry, and calibration of the LT, LCO, NTT, and SMARTS imaging to the Sloan Digital Sky Survey \citep[SDSS;][]{Eisenstein2011} and the American Association of Variable Star Observers Photometric All-Sky Survey (APASS) standard stars in the field was achieved using the {\sc iraf daophot} package and a custom {\sc python} script.
NOTCam reductions (differential flat-fielding, sky subtraction, distortion correction and stacking of dithered images) were done using a modified version of an external IRAF package notcam (v 2.5)\footnote{http://www.not.iac.es/instruments/notcam/ guide/observe.html\#reductions}. $JHK$ bands were calibrated to Two Micron All-Sky Survey \citep[2MASS;][]{Skrutskie2006} sources in the field.
GROND images were reduced by the GROND pipeline \citep{Kruhler2008}, which applies de-bias and flat-field corrections, stacks images, and provides astrometric calibration.
NTT spectra were reduced via a custom pipeline as described in \citealt{Smartt2015}. 
SPRAT spectra were reduced and calibrated in wavelength via the pipeline described in \citet{Piascik2014} and \citet{Barnsley2012}, and calibrated in flux via a custom {\sc python} pipeline. 
WiFeS spectral reduction is described in \cite{Childress2016}.

 \begin{figure}
 	\centering
 	\includegraphics[scale=0.45]{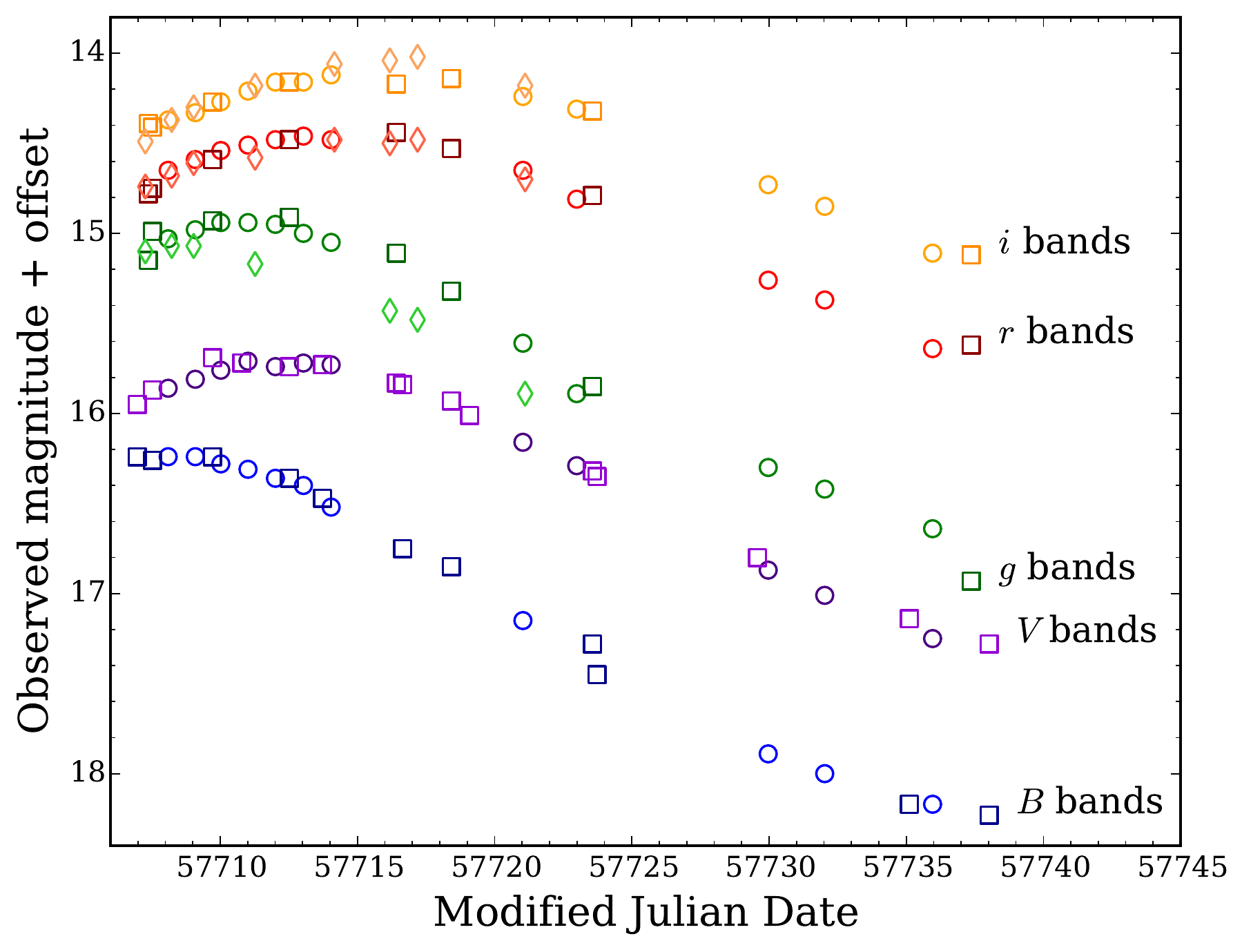}
 	\caption{A comparison of the effects of different filter responses on the early-time light curve of SN 2016iae; LT (circles), GROND (diamonds), LCO (squares).}
 	\label{fig:filters}
 \end{figure}
 
The filters used on the LT, GROND, and LCO telescopes are all subtly different, as demonstrated in Figure~\ref{fig:filters}. For the most part effects are small but most pronounced for the GROND~$g'$ filter ($\lambda_\mathrm{eff}\sim 4505$ \AA) which is bluer than the IO:O SDSS-$g$ filter ($\lambda_\mathrm{eff} \sim 4696$ \AA) and so probes an area of the spectrum closer to the $B$-band ($\lambda_\mathrm{eff} \sim 4348$ \AA). This leads to a clear difference in the light curves of GROND-$g'$ and the other $g$ filters. Each band is treated independently throughout.

\subsection{Host galaxy distances and reddening}
Distances are taken from the NASA Extragalactic Database\footnote{https://ned.ipac.caltech.edu/} (NED) and assume a cosmology of $H_0=73$ km s$^{-1}$ Mpc $^{-1}$, $\Omega_m=0.27$, $\Omega_\Lambda=0.73$ throughout. 
Distances to objects outside of the Hubble flow can be highly uncertain, however, our error calculations carry no uncertainty for $\mu$. This is to allow the photometric properties calculated in this work, and their associated errors, to be easily projected to different distances. 

Galactic reddening \Emw\ for each object is found in \citet{Schlafly2011}. 
Host reddening \Eh\ is estimated using the equivalent width of the host-galaxy \NaI\ D doublet absorption feature in the spectra \citep{Poznanski2012}, although there are limitations to this method when applied to low resolution spectra \citep{Poznanski2011}. 
The equivalent width of the \NaI\ D line is calculated in relation to a ``local continuum'', defined as the boundaries of the \NaI\ absorption line.
While it is found that S/N and the estimated local continuum affect the value of the equivalent width, the largest contribution to the error is the uncertainties given in \cite{Poznanski2012}. 
This method also assumes a Milky Way type extinction law with $R_V = 3.1$, which does not necessarily apply to other galaxies \citep[e.g., M82;][]{Ashall2014}. Also the equations used to derive \Eh\ are very sensitive to the equivalent width and when the lines saturate the relationship is no longer valid.
Other methods of deriving \Eh\ use colours to estimate the reddening \citep{Drout2011,Stritzinger2018}. These methods are not used here, but tests of various SNe show that they broadly agree.

%%%%%%%%%%%%%%%%%%%%%%%%%%%%%%%%%%%%%%%%%%%%%

%% Light curves

%%%%%%%%%%%%%%%%%%%%%%%%%%%%%%%%%%%%%%%%%%%%%
\section{Light curves}\label{sec:lcs}
\subsection{Multi-colour light curves}

\begin{figure*}
	\centering
	\includegraphics[scale=0.55]{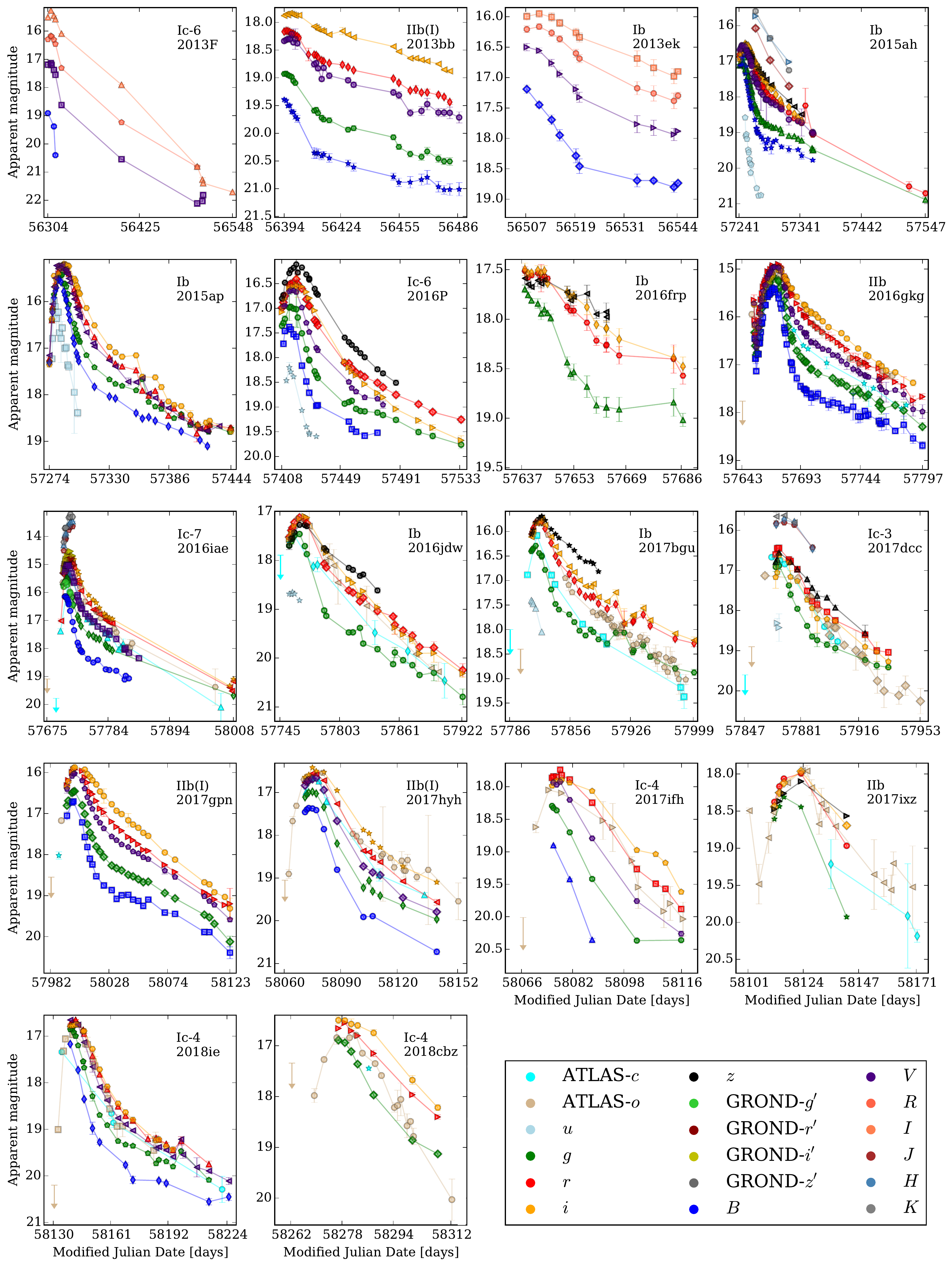}
	\caption{The observed multi-colour light curves of the SNe in the sample. For clarity, similar bands (e.g., $g$ and $g_p$) have been grouped together despite subtle differences in filter response functions. }
	\label{fig:lcs}
\end{figure*}

The multi-colour light curves for each SN are shown in Figure~\ref{fig:lcs}, with the photometry listed in Table~\ref{tab:phottab}.

The light curve of type IIb SN 2016gkg \citep{Tartaglia2017,Arcavi2017,Bersten2018} shows a very clear shock-cooling tail, which is characteristic of many SNe IIb with extended envelopes \citep[e.g., SN 1993J, SN 2011fu, SN 2013df;][]{Richmond1994,Kumar2013,MG2014}. The ATLAS light curve of the type IIb SN 2017ixz suggests that a shock cooling tail is detected, with a 1 mag decline in 4 days before the light curve rises (there is a robust non-detection at $o$ = 20 mag just two days before the first point). We have no indication of shock cooling in other H-rich SNe, although the data sampling is insufficient to constrain this quantitatively.

\begin{table}
    \centering
    \caption{Table of photometry.}
    \begin{tabular}{lcccc}
        \hline
        SN & MJD & band & mag & Facility \\
        \hline
        2013F & 56304.08 & B & 18.91$\pm{0.05}$ & NTT \\
        2013F & 56312.04 & B & 19.38$\pm{0.05}$ & NTT \\
2013F & 56314.05 & B & 20.40$\pm{0.05}$ & NTT \\
2013F & 56304.08 & V & 17.19$\pm{0.05}$ & NTT \\
2013F & 56312.04 & V & 17.37$\pm{0.02}$ & NTT \\
 
        \hline
        \multicolumn{5}{l}{The full table can be found in machine readable format online.}\\
    \end{tabular}
    
    \label{tab:phottab}
\end{table}

%%%%%%%%%% BOLOMETRIC LCs %%%%%%%%%%%%%%%%%%
%%%%%%%%%%%%%%%%%%%%%%%%%%%%%%%%%%%%%%%%%%%%%%%%%%%

\subsection{Bolometric light curves}
\begin{figure*}
	\centering
	\includegraphics[scale=0.5]{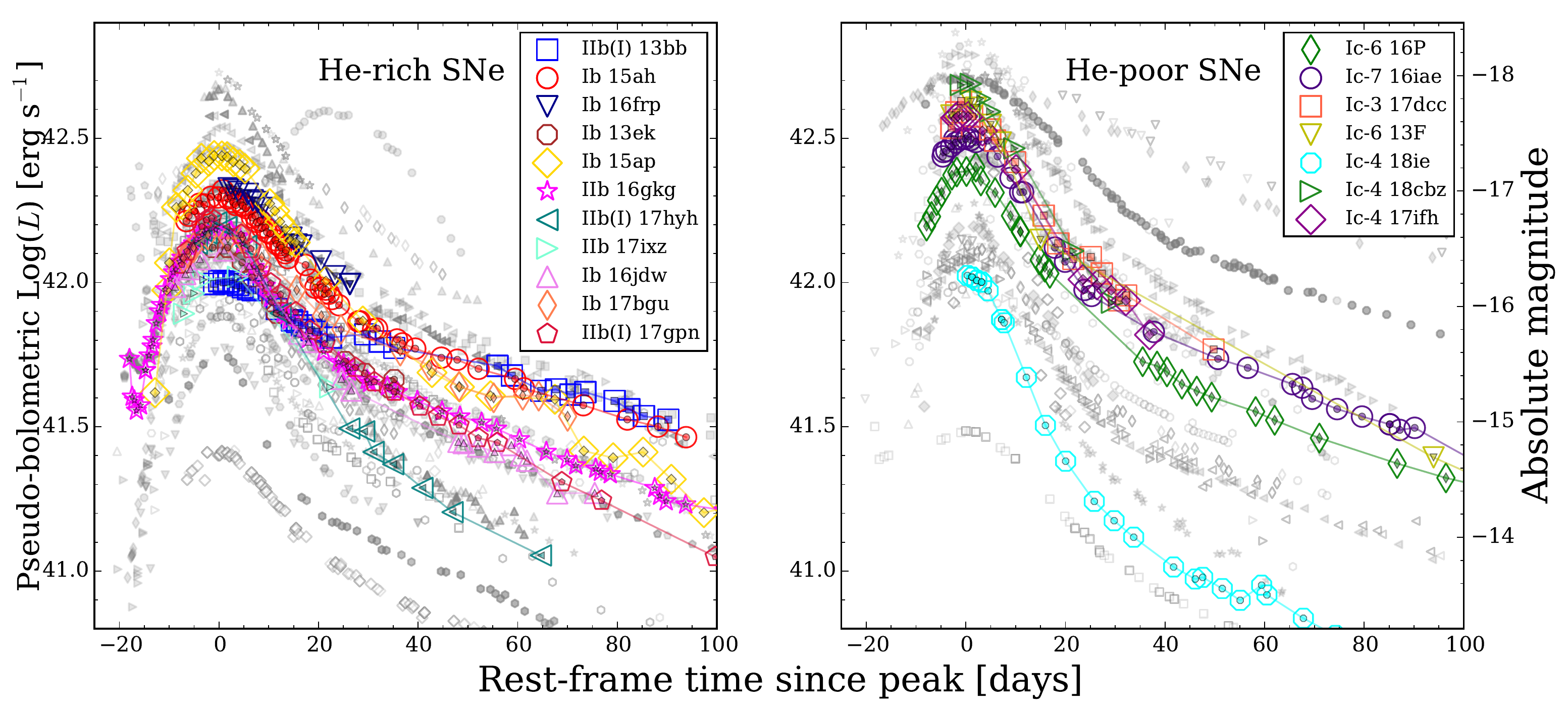}
	\caption{The 4000 -- 10000 \AA\ pseudo-bolometric light curves for (left) He-rich SNe and (right) He-poor SNe compared with the sample of P16 (grey). Note the shallow light curve of SN 2013bb. Grey open markers denote SNe for which the host extinction is unknown.}
	\label{fig:bols}
\end{figure*}

Optical 4000 -- 10000 \AA\ (pseudo-)bolometric light curves are constructed by converting the de-reddened apparent magnitudes to monochromatic fluxes, using the extinction law given in \cite{CCM} and zero-points from \citet{Fukugita1996}, and then integrating the resulting spectral energy distribution over the wavelength range.
The bolometric flux is then converted to bolometric luminosity using the distance modulus given in Table~\ref{tab:SNe}. The resulting light curves are shown in Figure~\ref{fig:bols} and compared with the sample from \cite{Prentice2016}, hereafter P16.

\subsubsection{Measuring physical parameters}
Light curve properties are derived using a low order (3-5) spline from {\sc univariatespline}, part of {\sc scipy}. 
The peak luminosity \lp\ is measured, plus three characteristic time-scales; 
\trise, \tdecay, and \dlate. 

The time-scales \trise\ and \tdecay\ are the times taken for the light curve to rise from \lp/2 to \lp\ and decay from \lp\ to \lp/2 respectively.
Uncertainties for these two values incorporate the errors on the spline fit, the photometric calibration, and \Etot. 
The linear decay rate at 100 days, \dlate, in units of mag d$^{-1}$, is measured via a linear fit through the light curve over several tens of days around, or as close as possible to, 100 days past maximum when the SN is on the \Cofs\ tail.\footnote{$\tau_{1/2}($\Cofs$)= 77.27$ days}
If the quality of the photometric data is good then the error on this value is negligible as given by the standard deviation of the covariance matrix used in the least squares fitting. The goodness of fit is also tested by changing the range of time over which the function was applied, however it was found that the maximum deviation was a thousandth of a magnitude per day and typically lower.
At this phase, the mass of $^{56}$Ni synthesised during explosive silicon burning, and the deposition energy from the decay products (positrons and $\gamma$-rays) of $^{56}$Co into the optically thin ejecta is probed. 
Full trapping of $\gamma$-rays would result in a decline of $0.0098$ mag d$^{-1}$ and a decline rate greater than this indicates less efficient trapping of the \Cofs\ decay products.
On the other hand, a slower decay rate suggests some different powering source \citep[as was the case for iPTF15dtg;][]{Taddia2018c}
All temporal parameters are measured in the supernova rest-frame. 
The results are given in Table~\ref{tab:bolstats} and discussed further in Section~\ref{sec:comp}.

\subsubsection{Light curve modelling, \Nifs\ mass, and rise time}
The time taken for the light curve to rise from explosion to maximum light \tp\ is used to derive the \Nifs\ mass (\mni) synthesised during the explosion, but
unfortunately this value is seldom known accurately. 
Estimates can be made by fitting a simple quadratic to the SN light curve on the rise but this method requires a well sampled rise. 
Alternatively a robust non-detection shortly before discovery can help constrain the explosion date.

To overcome this limitation and estimate \tp\ for all SNe, the pseudo-bolometric light curves are fit with the analytical model of \cite{Arnett1982}, as formulated for the photospheric phase in \cite{Valenti2008b}.
This model assumes homologous expansion, spherical symmetry, constant opacity, full $\gamma$-ray trapping, and that \Nifs\ is located centrally.
The light curve model was fit to the data using a Monte-Carlo routine that allowed the time of explosion \texp, the characteristic time scale of the light curve model \taum, and \mni\ to vary. 
The explosion time \texp\ was not allowed to take a value that is after the date of discovery, and nor was the model light curve allowed to exceed the limit set by any earlier non-detection. Information on discovery dates and corresponding photometry was found via Bright Supernova\footnote{http://www.rochesterastronomy.org/snimages/} \citep{GalYam2013b}.

The model fits were found to be reasonable but not perfect, and the relationship between \taum\ and \tp\ is complex. 
The ``best fit'' models to early data tended to underestimate the rise time indicated by the underlying data, by a couple of days. 
By forcing the model to include more post-max data points the peak of the model could be matched with the peak of the data, at the expense of the quality of fit.
Given this uncertainty, we make the assumption that \taum\ and \tp\ are approximately equivalent, with the value being the average of the best fit and forced peak fit to the data. 
Finally, $\delta$\mni\ is calculated from $\delta$\lp\ and $\delta$\tp\ using the formulation in \cite{Stritzinger2005}.
We find \mni\ values in the range 0.026 -- 0.19 \msun, with \tp\ between 7.8 -- 20.7 d (see Table~\ref{tab:bolstats}). \mni\ is discussed in relation to other events in Section~\ref{sec:nifsdebate}.

\subsubsection{Estimating ejecta mass}\label{sec:massmethod}
The ejecta mass \mej\ of the supernova is estimated through the following relation:
\begin{equation}
$\mej$ = \frac{1}{2}\left( \frac{\beta c}{\kappa} \right)$\taum$^{2} v_\mathrm{sc}
\label{eqn:arnett}
\end{equation}
Here $\beta \approx 13.7$ is a constant of integration, $\kappa$ is the opacity of the ejecta, and $v_\mathrm{sc}$ is a ``characteristic'' velocity of the ejecta. 
When applied to this model, $\kappa$ is often taken to be constant with a value between $0.07-0.1$ cm$^{2}$ g$^{-1}$, but in reality is variable throughout the SN ejecta \citep[see][]{Nagy2018}.
Additionally, defining a ``typical'' velocity in the form of \vsc\ is difficult, as the ejecta has a continuous velocity profile.
\vsc\ is commonly taken to be the photospheric velocity, \vph, at maximum light or some estimated expansion velocity. 
\mej\ is estimated here by using \taum\ $=$ \tp, assuming that $\kappa=0.07$ cm$^{2}$ g$^{-1}$ \citep[see][]{Taddia2018}, and that the scale velocity is given by the lowest measurable line velocity at maximum light.
In the case of He-poor SNe, this is \SiII\ \lam\ 6355, while for He-rich SNe it is \OI\ \lam\ 7774 if present or \HeI\ \lam\ 5876 if not.

The resulting values for \tp, \mni, \vph, and \mej\ are given in Table~\ref{tab:bolstats}.
\mej\ is given as an absolute value because the true uncertainty associated with this is unquantifiable (we do not know how accurate -- or wrong --  a 1D model is). 
Equation~\ref{eqn:arnett} is most sensitive to the value of \tp, such that an uncertainty of $\pm2$ days leads to variations of $\sim \pm$30 \%\ in \mej, but the presence of an uncertainty in the calculation may lead to undue confidence in the result.
The results are discussed in the context of other SE-SNe in Section~\ref{sec:massdebate}. We note here that our derived ejecta masses are between 1.2 and 4.8 \msun\ and our sample contains the most massive H-rich SE-SNe to date; SN 2013bb.

\begin{table*}
	\centering
	\caption{Statistics derived from the 4000 -- 10000 \AA\ bolometric light curves}
	\begin{tabular}{lccccccccc}
    \hline
	SN	&	\loglp\	& \trise & \tdecay & \dlate$^{a}$  & \tp& \mni & \vph$($\tmax$)$ & \mej$^{b}$ \\
     & [\ergs] & [d] & [d] & [mag d$^{-1}$] & [d] & [\msun] & [\kms] & [\msun]  \\
     \hline
    2013F & $42.6\pm{0.2}$ & - & $11.5\pm{0.1}$ & - &$12\pm{1}$ & 0.15$\pm^{0.09}_{0.06}$& 9000& $1.4$\\     
     
	2013bb & $42.0\pm{0.1}$ & - & $56\pm{10}$ & 0.011(1) & $\sim25^{c}$ &$0.07\pm{0.02}$ & 7000 &$4.8$ \\
    
    2013ek & $42.13\pm{0.02}$ & - & $15.0\pm{0.7}$ & - &$13\pm{2}$ & 0.05$\pm{0.01}$ &6000& $1.2$\\
    
	2015ah & $42.29\pm{0.01}$ & - & $19.3\pm{0.2}$ &0.014(1) & $16\pm{1}$ & 0.092$\pm{0.007}$ & 7000 &$2.0$ \\
    
    2015ap & $42.50\pm{0.03}$ & $8.0\pm{0.2}$ & $14.2\pm{0.2}$ & 0.020(1) &$13.5\pm{1.0}$ & 0.12$\pm{0.02}$ & 9000&  $1.8$\\
    
	2016P & $42.39\pm{0.02}$  & $9.9\pm{0.4}$ & $13.9\pm{0.2}$ &0.014(1) & $14\pm{1}$ & $0.09\pm{0.02}$ & 7000  &$1.5$ \\
    
    2016frp & $42.32\pm{0.01}$ & $9\pm{1}$ & $21.6\pm{0.7}$ & - & $16\pm{1}$ & $0.08\pm{0.007}$ & 9000 & 2.5\\
    
    2016gkg & $42.19\pm{0.03}$ & $11.6\pm{0.6}$ & $20.4\pm{0.2}$ & 0.018(1) & $20.7\pm{0.5}$ & $0.085\pm{0.008}$ & 8000 & 3.7$^{d}$\\
    
	2016iae & $42.4\pm{0.2}$  & $8.2\pm{0.1}$ &  $15.2\pm{0.1}$ &0.018(1) & $15\pm{1}$ & $0.13\pm^{0.08}_{0.05}$ & 9000  &$2.2$ \\ 
    
    2016jdw & $42.08\pm{0.04}$ & $12\pm{2}$ & $17.1\pm{0.5}$ & 0.018(2) & 19$\pm{1}$ & 0.06$\pm{0.02}$ & 11000 & 4.3\\  
    
    2017bgu & $42.16\pm{0.02}$ & - & $22.5\pm{0.6}$ & 0.016(2) & 18$\pm{1}$ & 0.069$\pm{0.003}$ & 8500 & 2.9\\ 
    
    2017dcc & $42.6\pm{0.1}$  & $8\pm{1}$ &  $15.5\pm{0.5}$ & - & $15\pm{1}$ & 0.16$\pm{0.04}$ & 13000  &$3.1$ \\
    
    2017gpn & $42.23\pm{0.02}$ & $9.9\pm{0.1}$ & $17.3\pm{0.3}$ & 0.023(1) & 15$\pm{1}$ & 0.070$\pm{0.007}$ & 8000 & 2.0\\ 
    
    2017hyh & $42.20\pm{0.01}$  & $8\pm{1}$ &  $13.0\pm{0.1}$ & 0.029(1)   & $16\pm{1}$ & 0.069$\pm{0.005}$ & 11000  &$3.0$ \\
    
    2017ifh & $42.6\pm{0.1}$ & $9\pm{1}^{e}$ & $14.7\pm{0.3}$ & - & 13$\pm{2}$ & 0.16$\pm{0.01}$ & 16000 & 3.0\\
    
    2017ixz & $42.02\pm{0.02}$ & $9\pm{1}^{e}$ & - & - & $20.0\pm{0.5}$ & $0.056\pm{0.03}$ & 8000 & 3.5\\ 
    
    2018ie & $42.02\pm{0.01}$ & $5.0\pm{0.5}^{e}$ & $10.1\pm{0.5}$ & - & $7.8\pm{0.5}$ & $0.026\pm{0.002}$ & 23000 & 1.5\\
    
    2018cbz & $42.69\pm{0.03}$ & $8\pm{1}^{e}$ & - & - & $14\pm{2}$ & $0.19\pm{0.04}$ & 14000 & 3.0\\
    \hline
    \multicolumn{9}{l}{$^{a}$Uncertainty of last digit given in parentheses. $^{b}$See text for a discussion on the uncertainty in \mej.} \\
    \multicolumn{9}{l}{$^{c}$From Karamehmetoglu et al. (in prep.). $^{d}$\cite{Bersten2018} calculate \mej\ $\sim3.4$ \msun. $^{e}$Estimated from ATLAS photometry.}\\
	\end{tabular}		
	\label{tab:bolstats}
\end{table*}

%%%%%%%%%% TEMPERATURE EVOLUTION %%%%%%%%%%%%%%%%%%
%%%%%%%%%%%%%%%%%%%%%%%%%%%%%%%%%%%%%%%%%%%%%%%%%%%
\subsection{Temperature evolution}
\begin{figure*}
	\centering
	\includegraphics[scale=0.55]{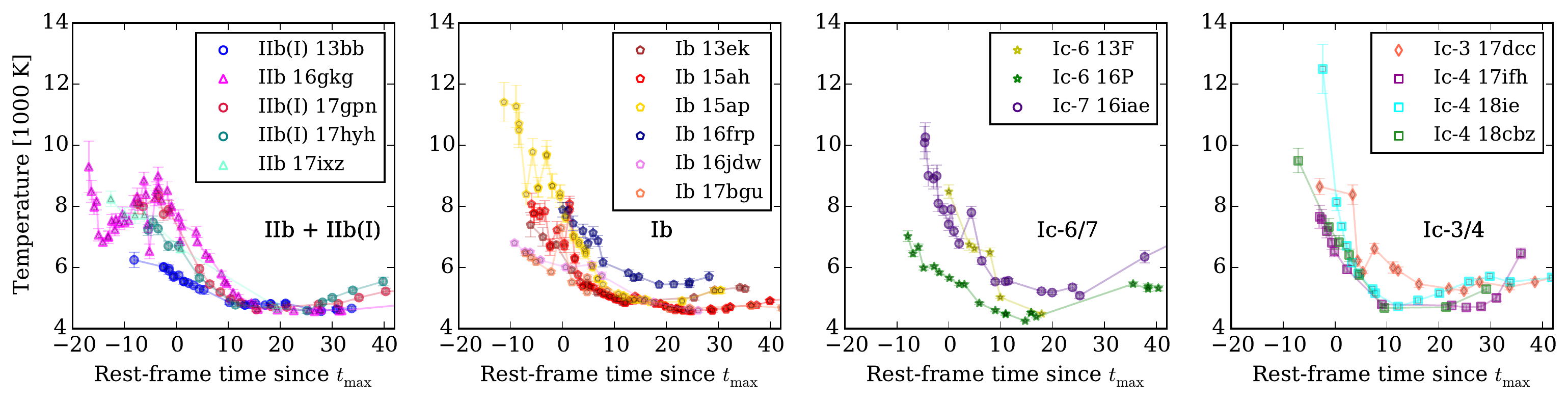}
	\caption{The temperature evolution of the SNe with respect to \tmax\ for H/He-rich SNe (top left), H-poor/He-rich SNe (top right), narrow lined He-poor SNe (lower left), and broad-lined SNe Ic (lower right).}
	\label{fig:T}
\end{figure*}

The temperature evolution of each SN during the photospheric phase was calculated by converting the de-reddened multi-colour photometry to monochromatic fluxes and constructing rest-frame SEDs.
A Monte-Carlo code was then used to fit the spectral energy distributions (SEDs) with a Planck function to find the best-fit temperature. Errors on the fit are calculated by allowing the flux at the beginning of each run randomly to take different values within the range of photometric errors. 
Variation due to uncertainty in \Etot\ was calculated and found to be within the MC-derived uncertainties, with the exception of SNe 2016iae and 2013F where $\delta$\Eh\ is large. Here the temperature errors can be $\pm25\%$ a week before maximum light, falling to $\pm8\%$ at +20 d.
Because uncertainty in \E\ does not act randomly (i.e., lower \E\ reduces all the temperature measurements) this contribution is not included in the plotted error bars.
Additionally, if spectra exist more than 1.5 days before the first photometric observation then a Planck function was fit to these to estimate the earliest temperature. For SN 2017ixz \citep{17ixzCLASS} and SN 2018ie \citep{18ieCLASS} this necessitated the use of classification spectra.

The temperature evolution of most SNe for the time observed is quite similar (Figure~\ref{fig:T}); an initial cooling phase with $T($\tmax$) = 4000 - 8000$ K, levelling off to $\sim 5000$ K around 7 -- 10 days after maximum light. 
The apparent rise in temperature after this is unlikely to be real however, as this epoch is defined by deviation from a thermal spectrum (i.e., emission lines begin to show) and a loss of the photosphere.

The H-rich SNe show a series of different behaviours. The most He-rich of these, SNe 2016gkg and 2017ixz, both had shock-cooling tails, and both had rise times of \tp\ $\sim 20$ d. 
The shock-cooling phase of SN 2016gkg is reflected in the temperature evolution showing a double-peaked profile, with the second peak occurring approximately 5 days before \tmax. SN 2017ixz does not show this behaviour, with the temperature $\sim 7$ d after \texp\ already being in decline. 
The type IIb(I) SN 2017gpn may have a secondary peak, but as the early phase of this SN was not observed in multi-colour photometry it is not possible to tell conclusively. The ATLAS-$o$ detection suggests there was no prominent shock-cooling tail.
Finally, the unusual SN 2013bb is rather cool in comparison, with $T\sim 6200$ K approximately 9 days before maximum light. 

The He-rich SN 2015ap was discovered early, allowing us to constrain the explosion date to within a couple of days. The first temperature estimate is approximately two days after explosion and is around 11000 K. The variation in the temperature curve may just reflect photometric errors, however the flattening at $\sim9000$ K between $-10$ and 0 d may be real (a black body fit to spectra taken $-5$ d gives $T\sim 8500$ K).
SN 2016frp remains hotter than the other SNe after \tmax, as is apparent at $+15$ d when the temperature of the SN is 1000 K higher than the rest.
SNe 2016jdw and 2017bgu have comparatively long rise times ($\sim$ 19 and 18 days respectively) and this is reflected in their temperature evolution. By the time of the first multi-colour observations they are already cooler and their temperature curves evolve more slowly than the other SNe Ib.

Of the SNe Ic-6/7, SN 2016P is apparently cooler than both SNe 2016iae and 2013F, and although the latter are affected by significant and uncertain \Eh, accounting for this uncertainty still places them above SN 2016P before $+10$ d. Its temperature evolution has more in common with type Ib SNe 2016jdw and 2017bgu.
A check against black body fits to the maximum light spectrum shows that $T=8000$ K fits the continuum well, but overestimates the location of the peak. A temperature of $T=6000$ K matches the peak of the spectrum (and the SEDs) but not the continuum.

For SNe Ic-3/4, attempts to calculate $T$ are complicated due to a combination of line blanketing and extended line-forming regions, which are seen as broad absorption regions. 
Here, flux in the near-UV is scattered to progressively redder wavelengths, finally escaping through ``windows'' of lower opacity which manifest themselves in the spectra as pseudo-emission peaks. 
As a result, the spectra and SEDs may not be well represented by a black body as the flux peak is not representative of $T$.
Examination of $T$ derived from SEDs and spectra for the ePESSTO SN in this subgroup, SN 2017dcc, reveals a similar issue to SN 2016P. $T$ estimates derived from SED fits and spectral continuum fits ($> 5000$ \AA) are in agreement, but these temperatures fail to match the shape of the spectrum in the blue. This would be consistent with flux redistribution however, and under this assumption, these SNe show similar temperature and evolution around maximum light. SN 2018ie is initially very hot some $\sim 3$ days after explosion (see ATLAS limits) and undergoes rapid cooling and remains comparatively hot after \tmax.
Note, that despite similar issues between SN 2016P and the Ic-3/4 SNe in terms of calculating $T$, SN 2016P is much cooler.
The classification of SN 2016P is ambiguous; it was originally classified as a ``Ic-BL'' \citep{Zhang2016ATEL}, but the spectra are heavily contaminated by host-galaxy emission lines. Its spectral properties (velocities, shape) better fit SNe Ic-6/7, yet in this instance it better matches the broad-lined SNe.
The temperature evolution of the SNe presented here is consistent with that found by \cite{Taddia2018}.

%%%%%%%%%%%%%%%%%%%%%%%%%%%%%%%%%%%%%%%%%%%%%

%% Spectra

%%%%%%%%%%%%%%%%%%%%%%%%%%%%%%%%%%%%%%%%%%%%%
\section{Spectroscopy}\label{sec:spectra}
The maximum light spectra of the 18 SNe are shown in the top panel of Figure~\ref{fig:spec}; plots of all spectra can be found in the supplementary material. Spectra not already released as part of a PESSTO data release will be uploaded to WISeREP\footnote{https://wiserep.weizmann.ac.il/} \citep{Yaron2012}.
Eight of the SNe were observed as they entered the nebular phase (Figure~\ref{fig:nebspec}). They show emission lines common to H-deficient core-collapse supernova, such as a series of \FeII\ lines at $\sim 5000$ \AA, \NaI\ D, \OI\ \lam\lam 7772, 7774, 7775 and the \CaII\ NIR triplet during the first month after explosion, and \Oneb\ \lam \lam\ 6300, 6363, [\CaII] \lam \lam\ 7292, 7324 during the nebular phase.

\begin{figure*}
	\centering
	\includegraphics[scale=0.55]{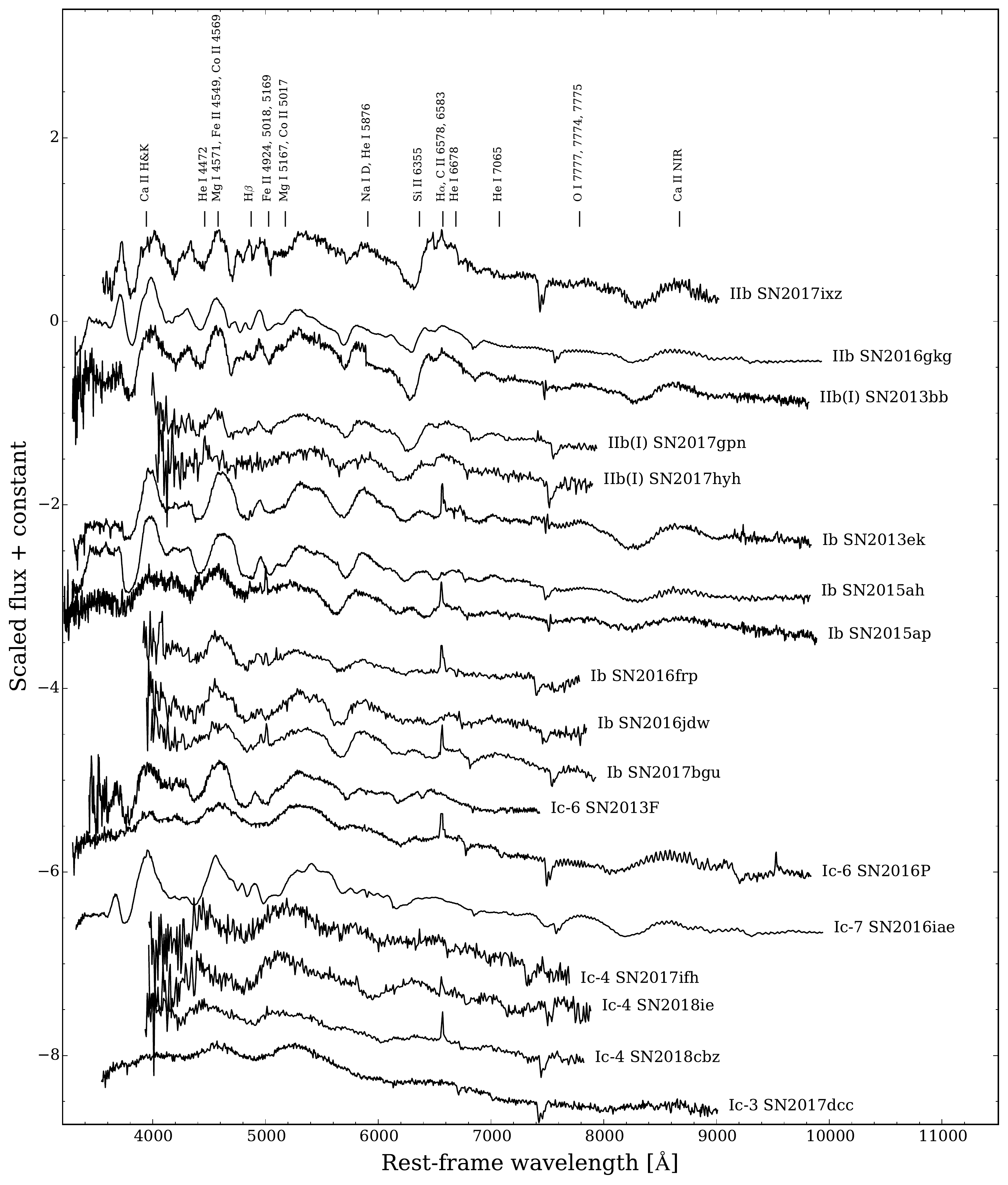}
	\caption{\tmax\ spectra of the 18 SE-SNe grouped according to sub-type. Prominent galaxy lines have been truncated in SNe 2016frp and 2016P, these greatly affect the appearance of the latter around 5000 \AA. Plots of all the spectra can be found in the supplementary material.}
	\label{fig:spec}
\end{figure*}

\begin{figure}
	\centering
    \includegraphics[scale=0.3]{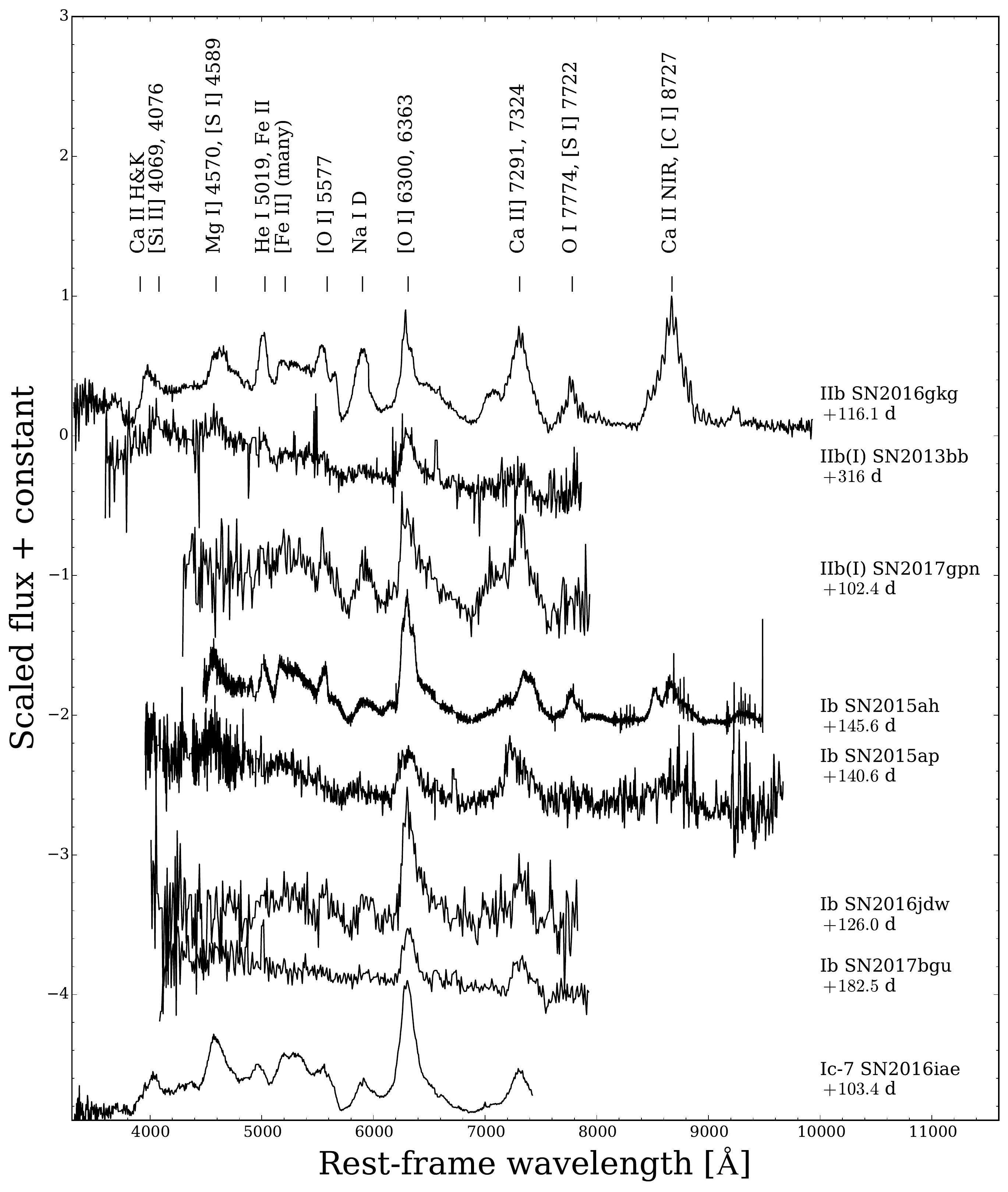}
	\caption{Nebular phase spectra of 8 SE-SNe, where galaxy lines have been truncated. Common emission lines associated with the nebular phase of core-collapse SNe are shown. Notice that even in H-rich SE-SNe there is no clear \Ha\ emission during this phase. }
	\label{fig:nebspec}
\end{figure}

\subsection{Measuring line velocities}
For all SNe, line velocities of common elements (e.g. H, He, Si, Fe, O, Ca and Na) are calculated by measuring the minimum of the absorption profile in velocity-space relative to the rest position of the ion.
The error on this value is calculated by taking the width $\delta v$ in velocity space of the minimum, where $df_{\lambda}(v)/dv \sim 0$. 
It is important to note that for blended lines (e.g. those in SNe Ic-3 and 4) the absorption minimum does not necessarily reflect any single Doppler-shifted atomic transition \citep{Prentice2017}, and the uncertainty represents the entire width of the feature. This approach results in very conservative errors.

\begin{figure*}
	\centering
	\includegraphics[scale=0.55]{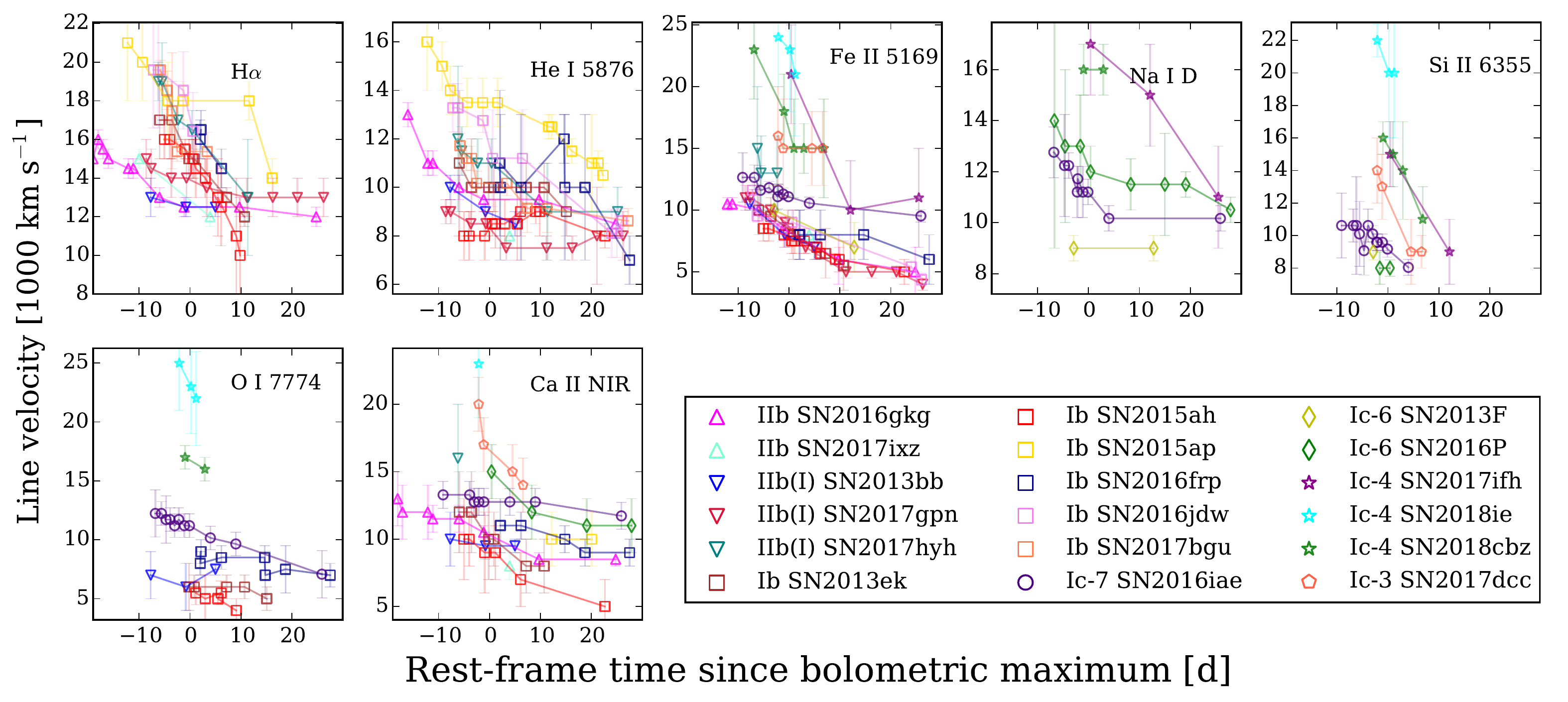}

	\caption{The photospheric phase velocity curves of common lines in SE-SNe.}
	\label{fig:vels}
\end{figure*}

The line velocities of the SN are measured during the photospheric phase ( $<30$ d after \tmax). For the H-rich SNe we measure the velocity as defined by the minima of the absorption profiles of \Ha, \HeI\ \lam\ 5876, \FeII\ \lam\ 5169, \OI\ \lam\ 7774, and the \CaII\ NIR triplet. 
For the He-poor SNe we measure \FeII, \OI, and \CaII\ as before, and further include \NaI\ D and \SiII\ \lam\ 6355.
The line velocity plots are shown in Figure~\ref{fig:vels}. It can be seen that Ic-4 SN 2018ie has the highest line velocities of any object in this sample.

\subsubsection{\Ha}
The \Ha\ velocities of the H-rich SNe show two contrasting behaviours. SNe 2016gkg, 2017gpn, 2017ixz and 2013bb show similar velocity evolution, at least where the data overlap in time. However, SN 2017hyh has high velocity \Ha\ that declines in velocity rapidly, by $\sim 6000$ \kms over 20 days and has a velocity $\sim 17000$ \kms\ at \tmax. At this epoch it is the highest velocity H-rich SE-SN known \citep{Liu2016,Prentice2017}.
Interestingly, this shows similarity to the \Ha\ velocity curves of the SNe Ib and those of \SiII\ in He-poor SNe. The \Ha\ line in SNe Ib declines rapidly to peak and usually disappears approximately a week later.
While the presence of H in SN 2017hyh is clear, the identification of this line in SNe Ib is ambiguous. 
It may be \SiII\ or a mix of this line with \Ha, but the behaviour here suggests that as the SN approaches peak the line is dominated by the \SiII\ component. 
The solution to this problem will require non-LTE spectral modelling.

\subsubsection{\HeI\ \lam\ 5876}
Both SNe 2015ap and 2016jdw are towards the upper end of the SN Ib \HeI\ \lam 5876 velocity distribution \citep[see][]{Liu2016,Prentice2017,Fremling2018}.
A curious behaviour is seen in the \HeI\ velocity of SN 2015ah which increases towards \tmax. This behaviour is unusual, as post-max increases in \HeI\ \lam 5876 velocities have been seen before (e.g., SN 2005bf) but never on the rise. Further analysis suggest that this only occurs in the  \HeI\ \lam\ 5876 line, and not in the other \HeI\ lines. This could be interpreted as the feature being a \NaI\ D/\HeI\ blend, and the evolution being due to changing line strength, however the feature is no stronger or broader than in other SNe Ib where this effect is not seen.

\subsubsection{\FeII\ \lam\ 5169}
The high velocity nature of SN 2017hyh is again apparent in the \FeII\ velocity curves, although post-maximum measurements are prevented by the poor S/N of the spectra.
Unfortunately no measurement of the \FeII\ velocity in SN 2015ap could be made due to contamination by host galaxy lines.
The velocity curve of SN Ic-7 2016iae is rather flat, as was noticed for Ic-7 SN 2007gr \citep{Valenti2008}, which is again further evidence for the ejecta of these SNe having a steep density profile.
The remainder of the SNe show velocity curves that decrease rapidly during the photospheric phase, and this occurs most rapidly in the Ic-3/4 SNe. 
These latter SNe have the highest line velocities here (but see the previous caveat on measuring velocities from broad lines).

\subsubsection{\NaI\ D}
The only velocity that could clearly be measured for Ic-6 SN 2016P is \NaI\ D. It shows a velocity curve similar in shape to Ic-7 SN 2016iae but at a higher velocity. As a contrast, the two measured velocities for Ic-6 SN 2013F are much lower, which was also reflected in the \FeII\ velocities.

\subsubsection{\SiII\ \lam\ 6355}
As with \NaI, \SiII\ is measured in He-poor SNe only. The behaviour of the velocity curves was discussed in relation to \Ha. However, SN 2016iae and SN 2013F again show the velocity difference seen in \NaI.
The measurements of SN 2016P are not robust, although if the line is present and not an artefact of the host emission lines then the velocities can be considered as upper limits. Clearly, this is hard to reconcile with the \NaI\ velocities.
The SN Ic-3/4 velocities are again much higher than for any other SN type and also decline far more rapidly.

\subsubsection{\OI\ \lam\ 7774}
The \OI\ line can be hard to measure due to a strong telluric feature that tends to lie at 8000 -- 10000 \kms\ from its rest wavelength in low-$z$ SNe.
The best sampled velocity curve is that of SN 2016iae which displays an almost linear decline. 
We also measure an exceptionally high velocity for SN 2018ie ($\sim 25000$ \kms).
The remaining measurements are hindered by large uncertainties but show that the \OI\ velocity in He-rich SNe varies significantly, more so than for \Ha\ or \HeI.

\subsubsection{\CaII\ NIR}
The \CaII\ NIR absorption feature can be easy to measure if the spectra extend sufficiently redward, but uncertainty in its measurement can come from the broad nature of the feature, and the wide wavelength separation of the triplet components (8498 \AA, 8542 \AA, and 8662 \AA), which means that the actual velocity is poorly constrained.
Velocities determined from \CaII\ NIR in SN 2015ah decrease rapidly, with no clear indication of levelling off (compare this with the other SNe). This behaviour was also present in all the other velocities that could be measured for SN 2015ah except for \HeI. This makes the evolution of the \HeI\ \lam\ 5876 line even more exceptional.
The velocity curve of SN 2016iae is again found to be flat, while the velocity of SN 2016P is high initially before falling below that of SN 2016iae.
Finally, a single measurement of SN 2017hyh confirms its high velocity nature in relation to He-rich SNe.

\subsection{Luminosity of \Oneb\ \lam\ 6300 and progenitor \mzams}
Oxygen is an effective coolant of the SN ejecta at late times. The luminosity of the \Oneb\ \lam\ 6300 line, $L_\mathrm{[OI]}$, was calculated by absolute calibration in flux of the nebular spectra to the available $R/r-$band photometry, correcting for \Emw\, then correcting for the effects of cosmological expansion, then correcting for \Eh.
Under the assumption that the emitting region is spherical, a single Gaussian profile was fit to the spectra centred on 6300 \AA\ and was allowed to extend to a width that remained inside the emission line.
This approach was taken to avoid including emission from other lines in the region.
A pseudo-continuum, defined from some featureless region close to 6000 \AA, was subtracted from the Gaussian flux. 
The Gaussian flux density was integrated over the wavelength region 6137 -- 6518 \AA\ (regions where $f_\mathrm{continuum}>f_\mathrm{Gaussian}$ were treated as zero) and the integrated flux converted to luminosity using the distance modulus. Errors are assessed by allowing the position of the pseudo-continuum to vary between $\pm{10}$ \AA\ from the reference position. This uncertainty was then added in quadrature to the uncertainty on the flux due to photometric errors.

\begin{figure}
	\centering
	\includegraphics[scale=0.43]{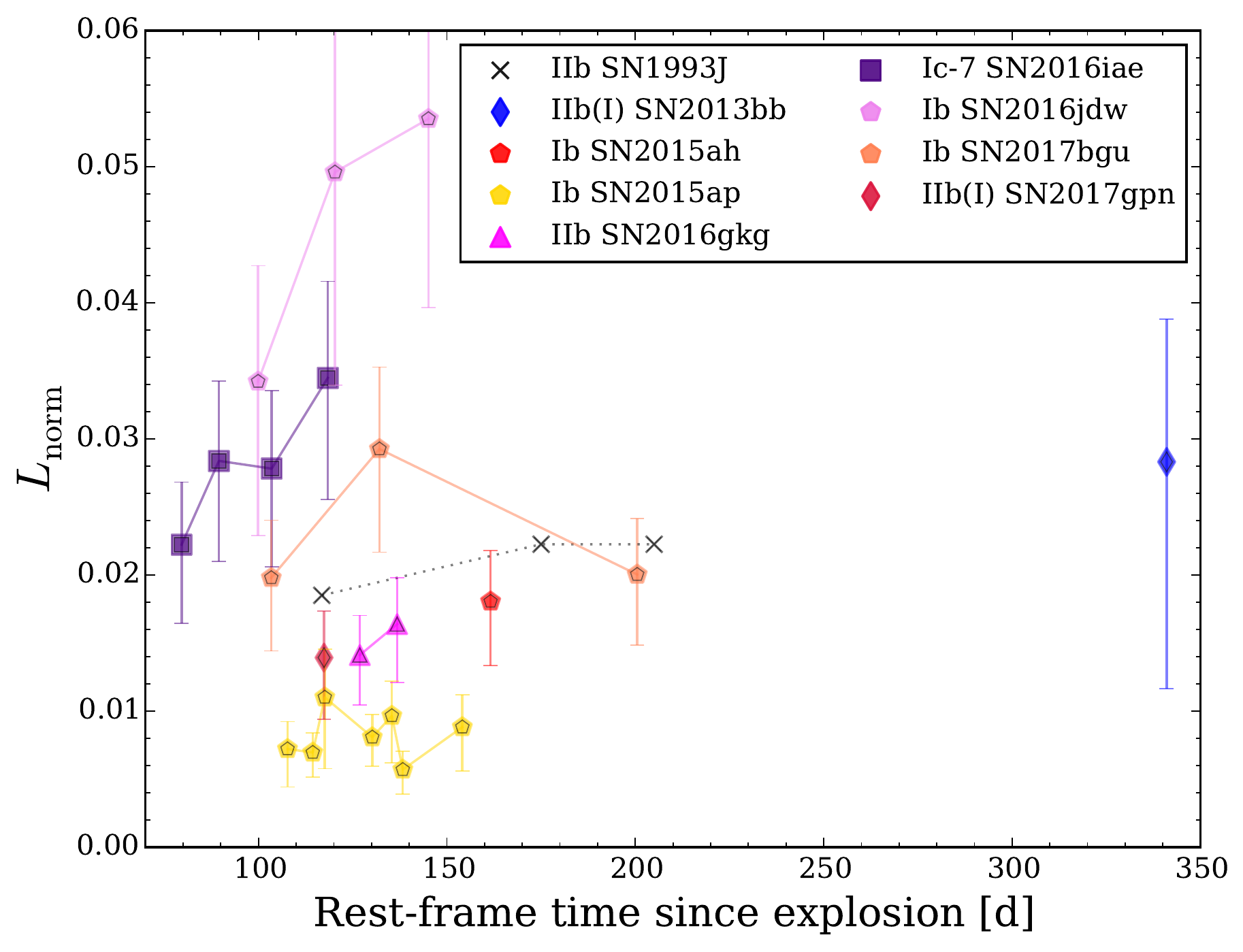}

	\caption{The normalised luminosity of the \Oneb\ \lam\lam\ 6300,6363 line as derived from the nebular and near nebular spectra. Shown for reference is $L_\mathrm{norm}$ of type IIb SN 1993J \citep{Barbon1995,Richmond1996}. }
	\label{fig:Olum}
\end{figure}

The flux is then normalised to the energy decay rate of \Cofs\ following the procedure of \cite{Jerkstrand2015}
\begin{equation}
    L_\mathrm{norm}(t) = \frac{L_\mathrm{[OI]6300}}{1.06\times 10^{42}\frac{M_\mathrm{Ni}}{0.075 M_\odot}   (e^{\frac{-t}{111.4d}}-e^{\frac{-t}{8.8d}})~\mathrm{erg~sec}^{-1} }
\end{equation}\label{eqn:Jerkstrand}
The results are shown in Figure~\ref{fig:Olum}, and we note that $L_\mathrm{norm}$ follows the behaviour of $L_\mathrm{[OI]}$. 
For Ic-7 SN 2016iae, $L_\mathrm{norm}$ increases between $\sim$120 and 160~d.
The oxygen luminosity of type Ib SN 2016jdw remains above that of the other SNe, by nearly 100\%\ in some cases. 
The late time photometry of SN 2013bb was estimated from EFOSC2 images to be $r\sim 21.5$ mag at $\sim$ 314 days after explosion.

 \cite{Jerkstrand2015} used $L_\mathrm{norm}$ to estimate the \mzams\ of the progenitor stars of several SNe IIb, under the assumption that it provides a good measurement of the He-core mass, which can be linked to \mzams\ \citep[this is model dependent, but to less than a factor of 2;][]{Jerkstrand2014}.
 They compared the $L_\mathrm{norm}$ curve with models ranging from \mzams\ $=12 - 17$ \msun, which contained oxygen masses of $M_\mathrm{O}= 0.3-1.3$ \msun.
 Comparison with their Figure 15 shows that SNe 2015ap and 2017gpn are within the range of models for which \mzams\ $=12-13$ \msun. 
 The progenitors of SNe 2015ah and 2016gkg lies between \mzams\ $=13-17$ \msun, but appear closer to the lower values. 
 The estimated initial mass of the progenitor of SN 2016gkg is within the range calculated in other works \citep[15-20 \msun;][]{Kilpatrick2017, Tartaglia2017} and sits at the same position as SN 2008ax in \cite{Jerkstrand2015}.
For SNe 2013bb, 2016iae, 2016jdw, and 2017bgu the progenitors may have initial masses around $17$ \msun. Their $L_\mathrm{norm}$ curves are close to the $17$ \msun\ model, and O production increases strongly with progenitor mass over 16 \msun\ \citep{Jerkstrand2015}.
Note that these values are based upon \mni\ calculated from the 4000 -- 10000 \AA\ bolometric light curve, which underestimates the peak luminosity by $\sim 30$ percent (P16) and would reduce the estimated \mzams.

%%%%%%%%%%%%%%%%%%%%%%%%%%%%%%%%%%%%%%%%%%%%%

%% Comparison with other SNe

%%%%%%%%%%%%%%%%%%%%%%%%%%%%%%%%%%%%%%%%%%%%%

\section{Comparison of physical properties with other SE-SNe}\label{sec:comp}

The properties of the 18 SNe can be compared with those found in larger samples.
In particular, we can examine how they fit within the distributions of $\mathrm{log}_{10}\left(L_\mathrm{p}\right)$ (hereafter \loglp), \trise, \tdecay, \dlate, \mni, and \mej.
Since the publication of P16 more usable data has become available, partially from public data accessed through the Open supernova Catalog\footnote{https://sne.space/} \citep[OSC;][]{Guillochon2017} and partially from recently published observations; 
SNe Ic-6 2006dn, IIb SN 2008aq \citep{Modjaz2014,Bianco2014,Stritzinger2018b}, Ic-4 SN 2016coi \citep{Prentice2018}, and GRB-SN 2016jca \citep{Ashall2017}.
These new SNe, analysed using the methods described in P16, provide an extension to that database and brings the total sample size to 106.\footnote{Not all of these have sufficient data coverage to be included in every distribution and GRB-SN 2011kl has been excluded as it was magnetar-powered \citep{Greiner2015}.}
Distributions are constructed using the available objects and the mean and median calculated. The standard deviation of each distribution is calculated and quoted with the mean, this represents the overall scatter of the objects.
Using the median, a 1 sigma region is calculated which includes 34 percent of the SNe above and below the median, thus enclosing 68 percent of the objects.

%%%%%%%%%%% LP distributions %%%%%%%%%%%%%%%%%%
%%%%%%%%%%%%%%%%%%%%%%%%%%%%%%%%%%%%%%%%%%%%%%%
\subsection{Distributions of bolometric peak luminosity }
\begin{figure}
	\centering
    \includegraphics[scale=0.42]{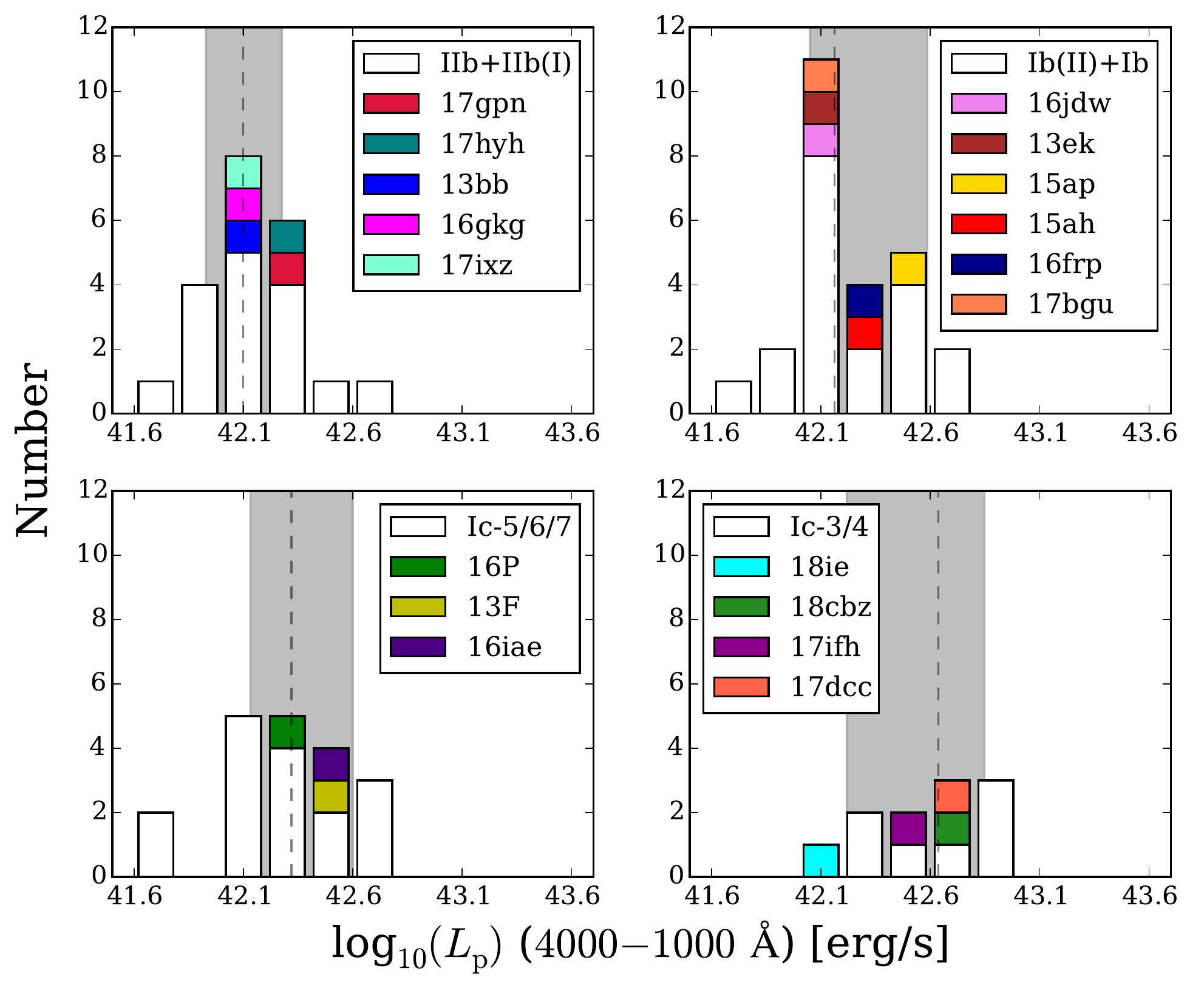}
	\caption{The distributions of \loglp\ with the new SNe in context, calculated from the 4000 -- 10000 \AA\ bolometric light curves, for SNe grouped by type. Only SNe with an estimated \Eh\ are included. The dotted line denotes the median of each distribution while $1\sigma$ around the median (enclosing 68\%\ of the objects in the distribution, 34\%\ either side of the median) is given by the grey regions. The distribution in white is SNe from the P16 sample. GRB-SNe are not included.}
	\label{fig:bolhists}
\end{figure}

The distributions of \loglp\ are shown by type in Figure~\ref{fig:bolhists}, with the statistics in Table~\ref{tab:lpstats}, where it can be seen that most of the SNe fall within one sigma of their respective medians as expected. 
The exception to this is SN 2018ie, which is underluminous compared with other SNe Ic-3/4. This cannot be attributed to underestimates in the distance or reddening.
As with previous results \citep{Drout2011,Taddia2015,Lyman2016,Prentice2016,Taddia2018} H-rich SNe are the least luminous on average, followed by H-poor/He-rich SNe, narrow lined SNe Ic, and broad lined SNe Ic. GRB-SNe are the most luminous SE-SNe.

\begin{table}
	\centering
    \caption{\loglp\ statistics}
    \begin{tabular}{lccc}
    \hline
    Type	&	Median & Mean & N \\
        &   [\ergs] & [\ergs] & \\
    \hline
    IIb + IIb(I) &  42.09$\pm{0.17}$  & 42.13$\pm{0.2}$    &  21  \\
    Ib + Ib(II) &   42.2$\pm^{0.4}_{0.1}$ & 42.3$\pm{0.2}$    & 25    \\
    Ic-5/6/7 & 42.3$\pm^{0.3}_{0.2}$   & 42.3$\pm{0.3}$    & 19    \\
    Ic-3/4 & 42.6$\pm^{0.2}_{0.4}$   & 42.6$\pm{0.3}$    & 11    \\ 
    XRF-SNe & 42.5$\pm{0.2}$ & 42.5$\pm{0.2}$ & 2 \\
    GRB-SNe & 42.96$\pm^{0.10}_{0.12}$ & 42.9$\pm{0.1}$ & 7 \\
    \hline
    \end{tabular}
	\label{tab:lpstats}
\end{table}

%%%%%%%%%%% trise and tdecay distributions %%%%%%%%
%%%%%%%%%%%%%%%%%%%%%%%%%%%%%%%%%%%%%%%%%%%%%%%%%%%
\subsection{The distributions of \trise\ and \tdecay}\label{sec:boltemps}

\begin{table}
	\centering
    \caption{\trise\ and \tdecay\ statistics per SN type}
    \begin{tabular}{lccc}
    \hline
    Type	&	Median & Mean & N \\
        &   [d]  &   [d]  &   \\
    \hline
    & \trise & & \\
    IIb + IIb(I) &  9.9$\pm^{0.8}_{1.9}$  & 10$\pm{1}$    &  14  \\
    Ib + Ib(II) &   10.4$\pm^{2.8}_{1.7}$ & 11$\pm{2}$    & 14    \\
    Ic-5/6/7 & 9.8$\pm^{3.3}_{3.1}$   & 10$\pm{2}$    & 16    \\
    Ic-3/4 & 8.6$\pm^{3.8}_{2.0}$   & 9$\pm{2}$    & 11    \\  
    XRF-SN & $\sim7$   & $\sim 7$    & 1    \\
    GRB-SNe & 9.7$\pm^{1.9}_{1.8}$   & 10$\pm{1}$    & 4    \\  
     \hline
    &  \tdecay & & \\
    IIb + IIb(I) &  15.1$\pm^{2.5}_{3.0}$  & 17$\pm{10}$    &  20  \\
    Ib + Ib(II) &   17.0$\pm^{4.7}_{3.4}$ & 18$\pm{5}$    & 29    \\
    Ic-5/6/7 & 17.5$\pm^{7.1}_{4.5}$   & 20$\pm{9}$    & 21    \\
    Ic-3/4 & 15.5$\pm^{5.1}_{2.6}$   & 16$\pm{4}$    & 11    \\ 
    XRF-SNe & 12$\pm{2}$   & 12$\pm{2}$    & 2    \\ 
    GRB-SNe & 14.3$\pm^{2.7}_{1.9}$   & 15$\pm{2}$    & 5    \\ 
    \hline
    \end{tabular}
	\label{tab:tstats}
\end{table}

\begin{figure}
	\centering
    \includegraphics[scale=0.34]{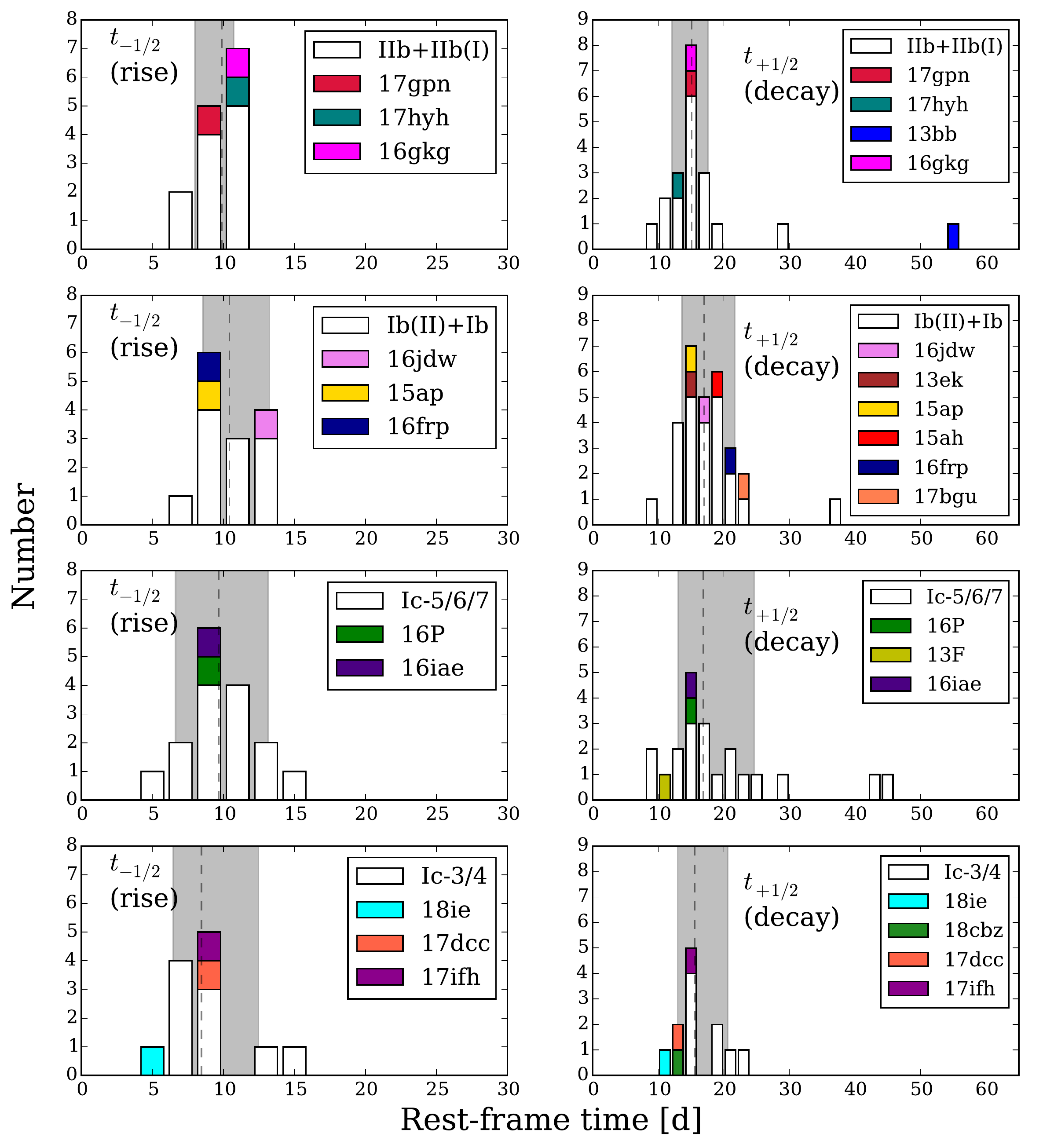}
	\caption{Distributions of \trise\ (left) and \tdecay\ (right) for the various SN subtypes. SN IIb(I) 2013bb is a clear outlier with \tdecay\ $\sim 56$ d. The dashed line and grey regions are as in Figure~\ref{fig:bolhists}. }
	\label{fig:tdist}
\end{figure}

The distributions for \trise\ and \tdecay\ are shown in Figure~\ref{fig:tdist} and the statistics in Table~\ref{tab:tstats}.
SN 2013bb is clearly an interesting object with \tdecay\ $\sim 56$ days, considerably longer than any other supernova in the sample.
the decay time of SN 2013F is short but highly uncertain due to lack of photometry.

The distribution of \tdecay\ is similar to that previously found but, compared with the smaller sample of P16, the differences in \trise\ found in that study between broad lined and non-broad lined SNe has largely vanished. The medians and means for these types in the present sample are within 0.6 d of each other. The Ic broad-line group retains its short average rise time, however there are some long-rise SNe in this group.

%%%%%%%%%%% trise against tdecay %%%%%%%%%%%%%%
%%%%%%%%%%%%%%%%%%%%%%%%%%%%%%%%%%%%%%%%%%%%%%%
\subsubsection{Comparison of \trise\ and \tdecay}
\begin{figure}
	\centering
	\includegraphics[scale=0.45]{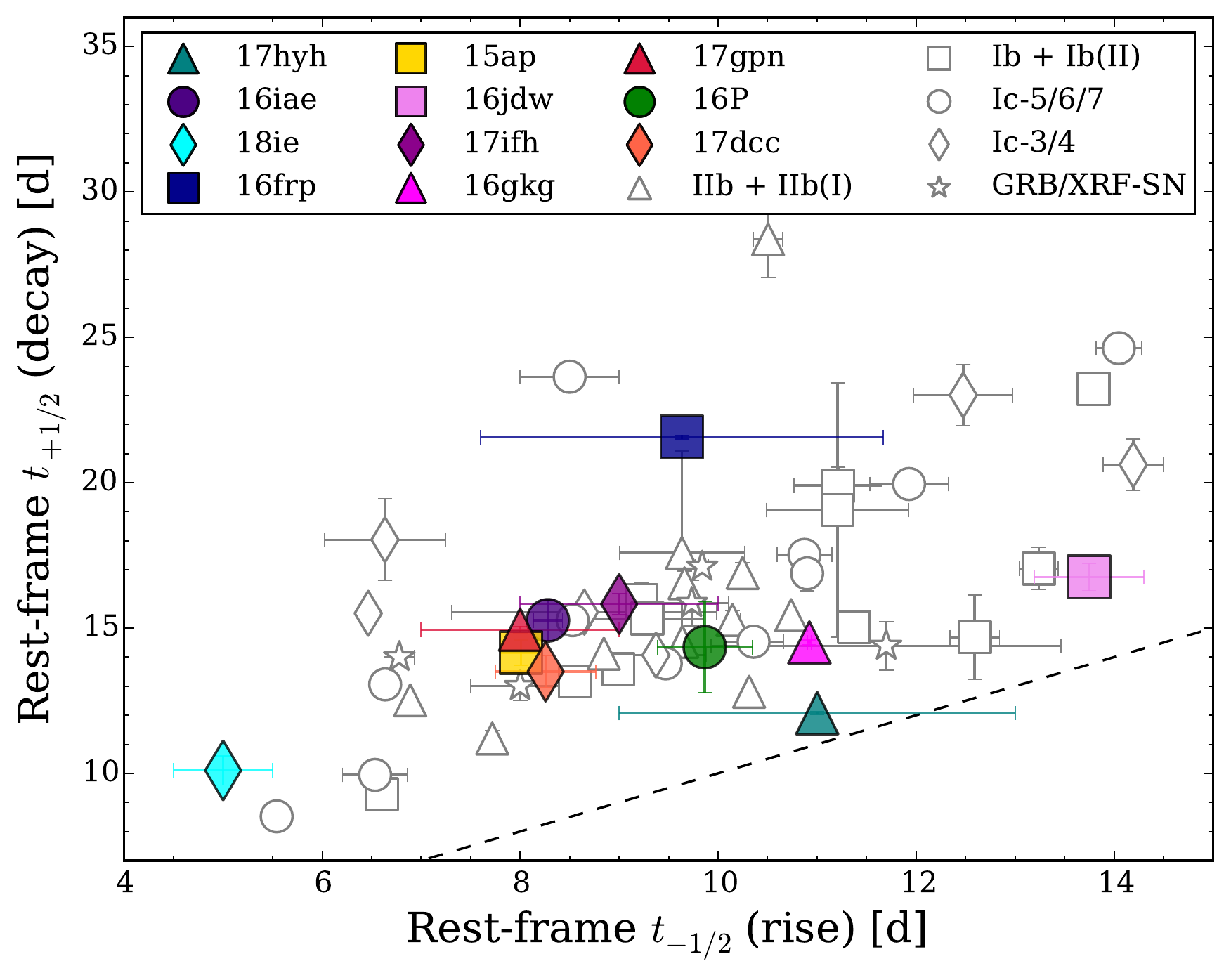}
	\caption{\trise\ against \tdecay\ for all SE-SN types. The markers denote the SN type, the dashed line is the line of equality, and the white symbols are the P16 sample.}
	\label{fig:risedecay}
\end{figure}

Figure~\ref{fig:risedecay} shows \tdecay\ as a function of \trise.  This plot is a measure of diffusion time and is affected by \mej, \ek, power source, and power source distribution. 
If the SN light curves are powered by different sources (e.g. \Nifs\ decay, interaction with the circumstellar medium or magnetar spin-down) then presenting the data in this way may reveal different sub-groups, (as attempted by, e.g.,  \cite{Nicholl2015} for superluminous SNe). 
For example, the early light curves of SNe IIP are not determined by radioactive heating \citep[see][]{Bersten2011}, so with \trise\ of a few days but \tdecay\ of many tens they would be placed in the far upper left of the plot. Likewise, rapidly evolving events \citep[e.g.,][]{Drout2014,Pursiainen2018,Prentice2018b} would be found in the extreme lower left hand corner.
The SE-SNe are clustered along a similar axis, which suggests that most have the same power source, in this case \Nifs\ decay. The outliers are the most interesting, as these could represent physical differences. SN 2018ie has both a fast rise and decline (with \tp$<9$ d) and a low luminosity, all of which the \Nifs-powered \cite{Arnett1982} model struggles to replicate. Such an investigation is outside the scope of this work however.

%%%%%%%%%%% dm100 distributions %%%%%%%%%%%%%%%
%%%%%%%%%%%%%%%%%%%%%%%%%%%%%%%%%%%%%%%%%%%%%%%
\subsection{The linear decay rate at 100 days}
\begin{figure}
	\centering
	\includegraphics[scale=0.42]{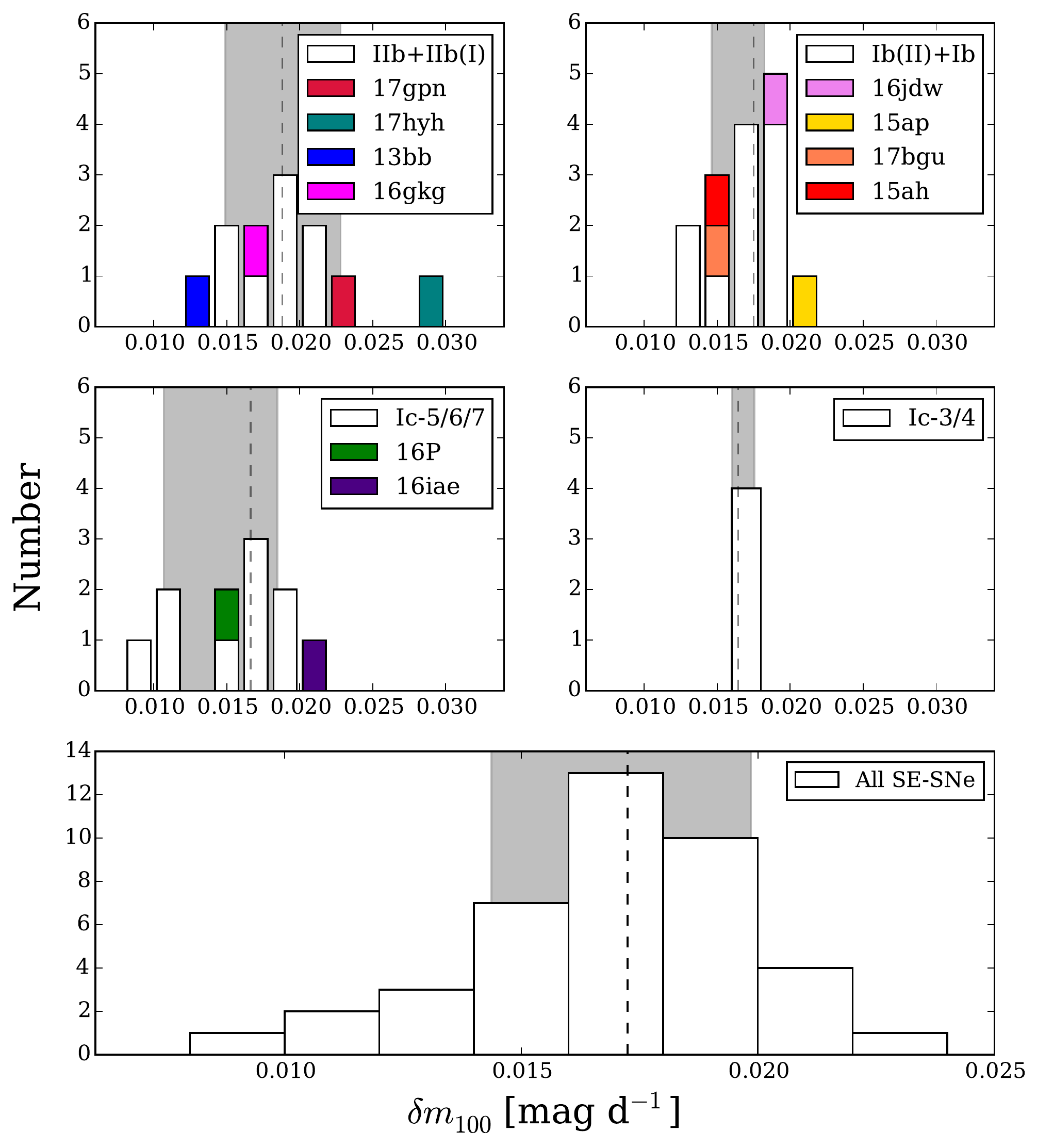}
	\caption{(top) The distribution of late time decay rates for SE-SNe; the median is $\sim$0.017 mag d$^{-1}$ for all SN types. Only SN 2013bb has had host-galaxy flux subtracted, but the effect of this on \dlate\ was found to be negligible at this phase, $\sim 0.001$ mag d$^{-1}$. The dashed line and grey regions are as in Figure~\ref{fig:bolhists}. }
	\label{fig:linhist}
\end{figure}

Figure~\ref{fig:linhist} (top) shows the distribution of \dlate, and it can be seen that SN 2013bb is one of the slowest declining SNe. By contrast, SN 2017hyh declines rapidly at $\sim 0.029$ mag d$^{-1}$. Most SNe decline faster, with the mean being $\sim 0.017$ mag d$^{-1}$ regardless of SN type. 

\begin{figure}
	\centering

    \includegraphics[scale=0.4]{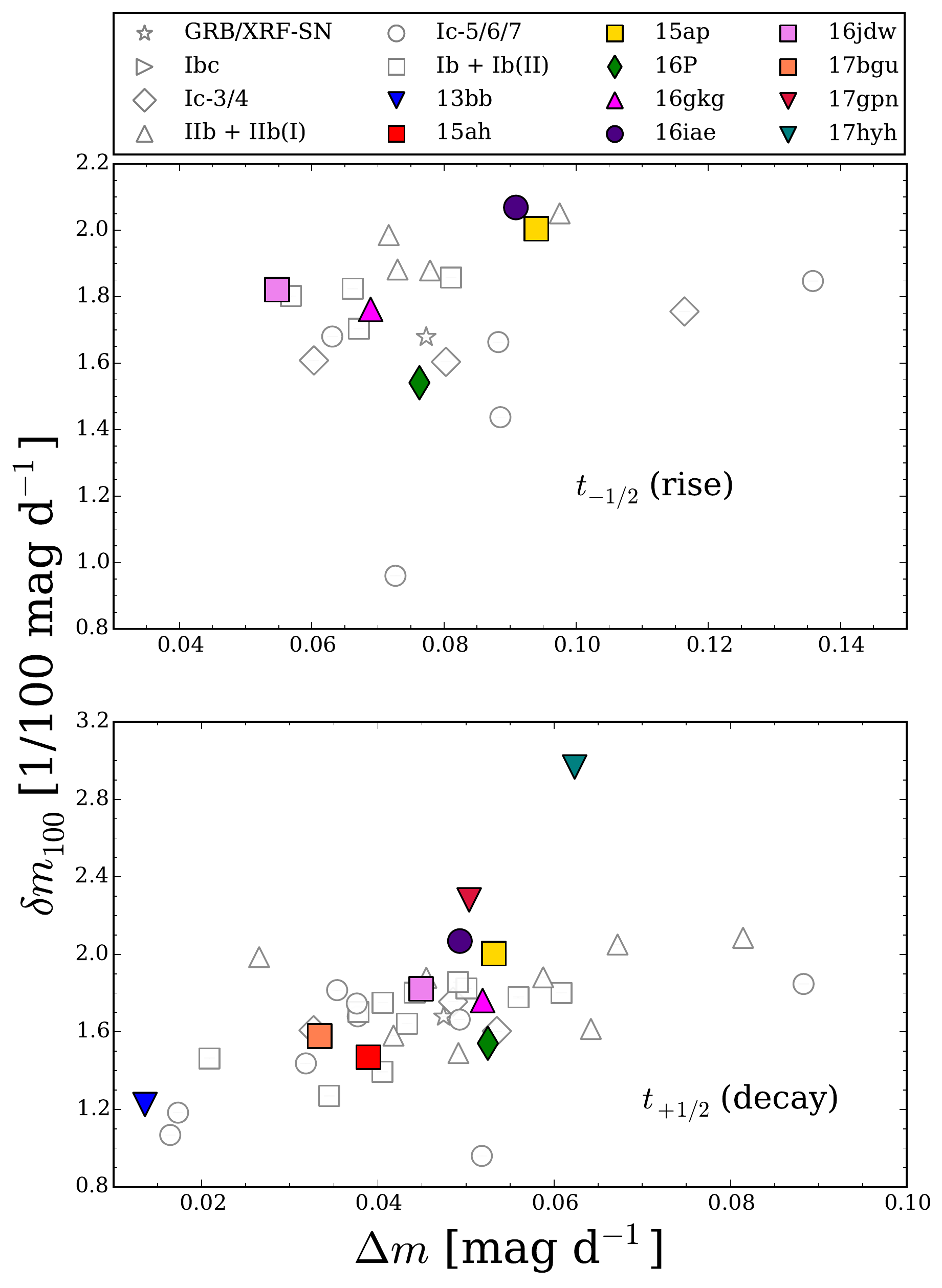}
	\caption{(Top) \dlate\ as a function of \trise\ in units of mag d$^{-1}$. (Bottom) \dlate\ as a function of \tdecay\ in units of mag d$^{-1}$. }
	\label{fig:dlate}
\end{figure}

Figure~\ref{fig:dlate} shows \dlate\ as a function of \trise\ and \tdecay, both of which are converted to units of \md. 
As when comparing \trise\ against \tdecay, both show a correlation in that a longer time-scale at early times corresponds to a longer time-scale at later times. This is  strongest for SNe Ib and SNe Ic-6/7 when considering \dlate\ against \tdecay.
In the bottom panel it can be seen that SN 2017hyh declines quickly by both metrics but its \dlate\ is the clearest outlier.

Finally, host-galaxy flux contaminates the SN flux at late times, when the SN is less luminous. This could have an impact on the values found here. However, we have one object with templates that allowed a host-subtracted bolometric light curve, SN 2013bb, and find that the host-subtracted \dlate\ increases over the non-subtracted version by just 0.001 mag d$^{-1}$. However, this effect may be more pronounced for brighter hosts, dimmer SNe, or SNe positioned over bright \HII\ regions, where conamination is more prevelant. 

%%%%%%%%%%%%%%%%%%%%%%%%%%%%%%%%%%%%%%%%%%%%%

%% Masses

%%%%%%%%%%%%%%%%%%%%%%%%%%%%%%%%%%%%%%%%%%%%%
\section{\Nifs\ and ejecta Masses}\label{sec:masses}
Analysis of the bolometric light curves allows for two masses to be calculated; \mni\ and \mej.
The \Nifs\ mass was calculated for a large sample of SNe in P16; here we add the calculations of the 18 SNe included here plus the 4 new literature SNe.
The ejecta mass is calculated for as many of the SNe in P16 as possible, using the line velocities from \cite{Prentice2017}, plus the new literature SNe, using the same method as described in Section~\ref{sec:massmethod}. A table of ejecta masses can be found in the supplementary material.
Here we consider the respective distributions for different subtypes.

\subsection{\mni\ distribution}\label{sec:nifsdebate}

\begin{table}
	\centering
    \caption{\mni\ statistics derived from 4000 -- 10000 \AA\ light curves}
    \begin{tabular}{lccc}
    \hline
    Type	&	Median & Mean & N \\
    & [\msun] & [\msun] & \\
    \hline
    IIb + IIb(I) &  0.07$\pm^{0.02}_{0.03}$  & 0.07$\pm{0.03}$    &  21  \\
    Ib + Ib(II) &   0.07$\pm^{0.10}_{0.02}$ & 0.09$\pm{0.06}$    & 25    \\
    Ic-5/6/7 & 0.09$\pm^{0.06}_{0.03}$   & 0.11$\pm{0.09}$    & 19    \\
    Ic-3/4 & 0.16$\pm^{0.06}_{0.10}$   & 0.15$\pm{0.07}$    & 11   \\
    XRF-SNe & 0.11$\pm{0.03}$   &  0.11$\pm{0.03}$   & 2    \\
    GRB-SNe & 0.30$\pm^{0.20}_{0.06}$   & 0.3$\pm{0.1}$    & 5   \\    
    \hline
    \end{tabular}
	\label{tab:nistats}
\end{table}

\begin{figure}
	\centering	\includegraphics[scale=0.42]{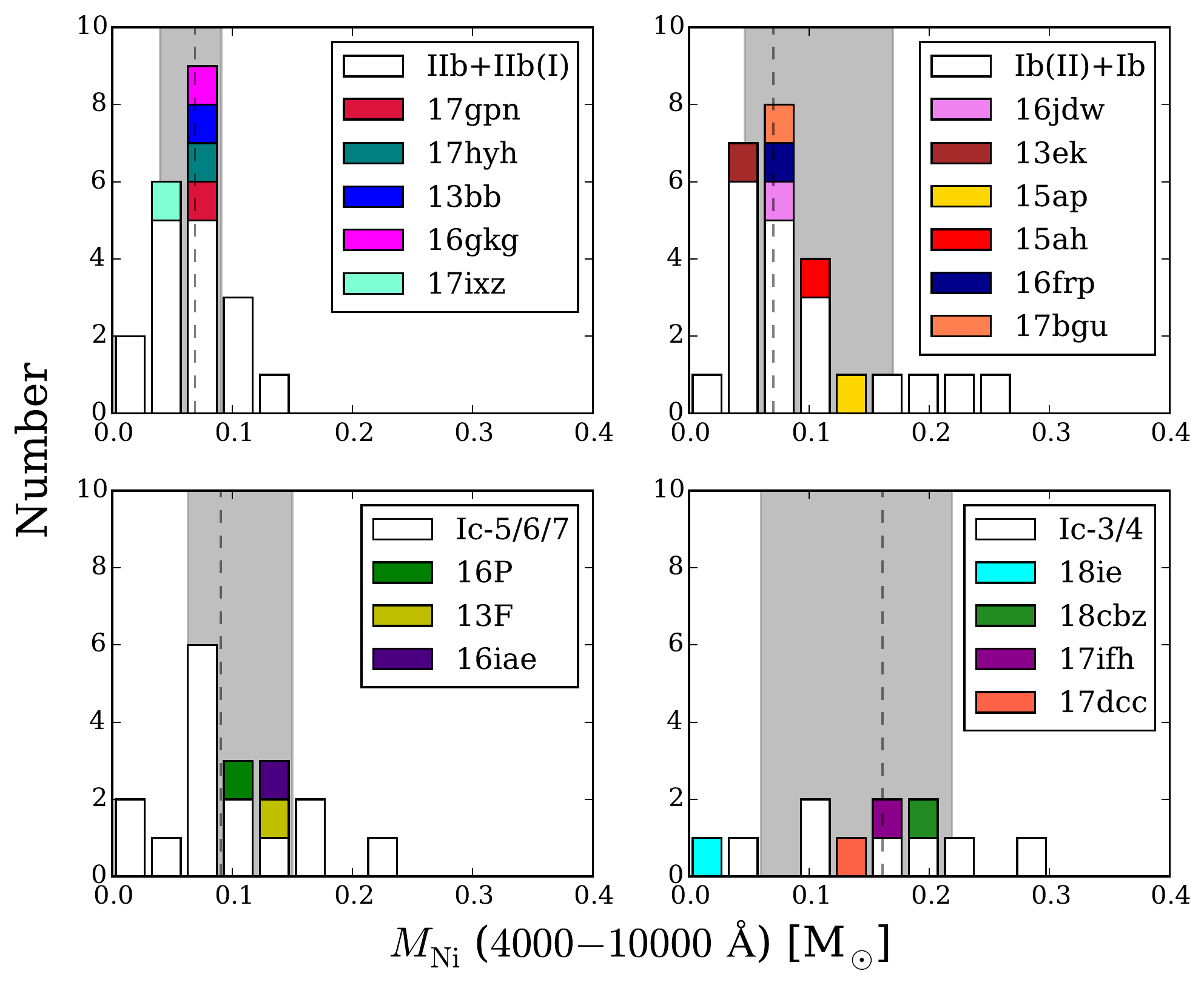}
	\caption{The distribution of \mni\ synthesised through explosive Si burning after core-collapse and calculated using the 4000 -- 10000 \AA\ light curve. Only SNe where \Eh\ is known are included. The dashed line and grey regions are as in Figure~\ref{fig:bolhists}. SNe Ibc have not been classified conclusively as either SNe Ic or SNe Ib.}
	\label{fig:MNihist}
\end{figure}

Figure~\ref{fig:MNihist} shows the distribution of \mni\ as derived from the $4000-10000$ \AA\ light curves for the different SN subtypes. The statistics for these distributions are listed in Table~\ref{tab:nistats}.  

Our 18 SE-SNe are not unusual in respect of their \mni.
The medians of distributions reflect the fact that energetic SNe synthesise the most \Nifs, and He-rich SNe the least.
Of the He-rich SNe, both H-rich and H-poor synthesise similar amounts of \Nifs\ if one considers the median. 
If one considers the mean then the He-poor explosions result in more \Nifs, which is in line with previous studies. 
Figure~\ref{fig:MNihist} shows that the \mni\ of the SNe Ib is more skewed than that of the H-rich SNe which is where the discrepancy in these two measurements lies. 
Consequently, we can say (for the most part) that the He-rich SNe produce similar masses of \Nifs\ but there exists a tail of higher \mni\ in the H-poor distribution.

\subsection{Ejecta masses}\label{sec:massdebate}

Analysis of ejecta mass for SE-SNe has also been done previously, but for smaller samples \citep[see][]{Drout2011,Taddia2015,Lyman2016,Taddia2018}. 
With the exception of \cite{Taddia2018}, where a hydrodynamics code was used, the procedure for determining \mej\ was similar to that used here, i.e. the method of \citet{Arnett1982}. In most cases an opacity parameter $\kappa$ of $0.05-0.07$ g cm$^{-2}$ was used, although \cite{Drout2011} assumed a typical value for \vph\ rather than using measurements, and \tp\ is often similar, but not identical, to that used here. These variations in input parameters can go some way to explaining differences in the results (see Table~\ref{tab:massstats}).
Additionally, differences between \mej\ calculated from photospheric phase spectral modelling and light curve modelling may arise for some SNe with long rise times and broad lines. For example, \mej\ of SN 1998bw differs between spectral modelling \citep[10 \msun;][]{Mazzali2013} and light curve fitting \cite[$\sim$5 \msun; e.g., this work,][]{Lyman2016}. The ejected mass is better constrained from nebular models however \citep[e.g.,][]{Mazzali2001,Mazzali2007,Mazzali2010}.

\begin{figure}
	\includegraphics[scale=0.42]{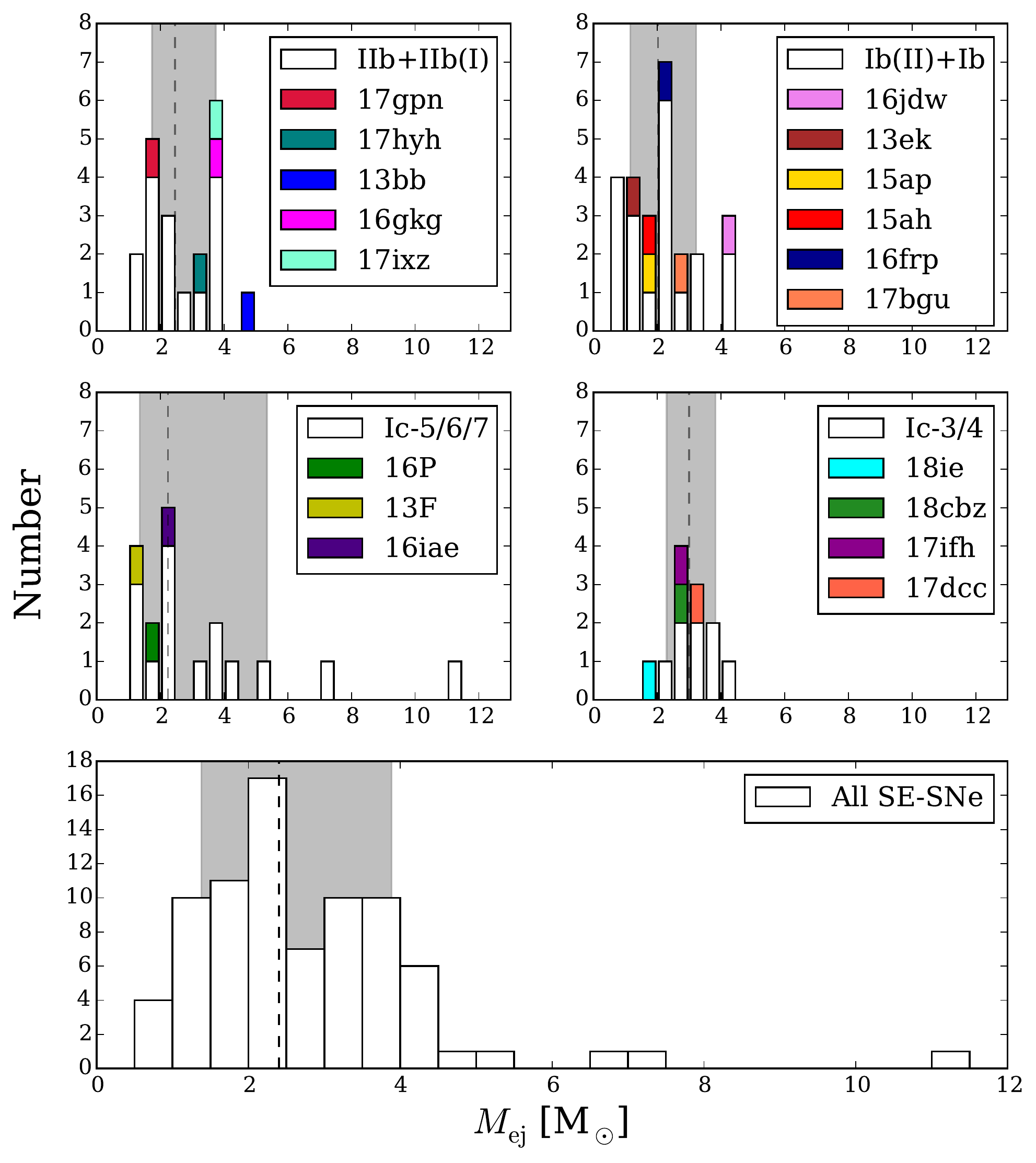}
	\caption{The \mej\ distribution of SNe IIb + IIb(I) (top left), SNe Ib(II) and Ib (top right), SNe Ic-5/6/7 (middle left), and SNe Ic-3/4 (middle right), and all SE-SNe including GRB and XRF-SNe (lower panel). The dashed line and grey regions are as in Figure~\ref{fig:bolhists}.}
	\label{fig:Icmej}
\end{figure}

Table~\ref{tab:bolstats} gives \mej\ for the SNe investigated here. 
Figure~\ref{fig:Icmej} shows the distribution of \mej\ for each subtype.
Our 18 SNe are, for the most part, typical in \mej\ but there are two clear exceptions. SN 2013bb is the highest mass H-rich SN in the sample with \mej\ $\sim 4.8$ \msun. This makes sense in the constext of its broad light curve and slow spectroscopic evolution.
Type Ib SN 2016jdw has \mej $\sim4.3$ \msun, and the only equivalent SNe Ib are SN 2009jf and the peculiar SN 2007uy with \mej\ $\sim 4.2$ \msun.

Most of the distributions are unimodal, with the exception of that for the H-rich SNe where there are two peaks around 1.9 and 3.9 \msun\ which appears to coincide with more compact and extended progenitors respectively (as determined by the luminosity and duration of the shock cooling tail). The distribution of SNe Ic-6/7 ejecta masses shows a large range of possible values which hints at a wide range of progenitor masses. 
The results here show that the median \mej\ is $<5$ \msun\ for all SNe. 
The bulk distribution for all SE-SNe has $<$\mej$>=2.8\pm{1.5}$ \msun\ (median \mej $=2.4$ \msun) and shows no clear indication of bimodality, but does have a ``high mass'' tail.

\begin{table}
	\centering
    \caption{\mej\ statistics in comparison with other work}
    \begin{tabular}{lccc}
    \hline
    Type$^{\dag}$	& Median & Mean & N \\
          & [\msun] & [\msun] &   \\
    \hline   
    & This work   & & \\
    IIb + IIb(I) & 2.5$\pm^{1.3}_{0.7}$	& 2.7$\pm{1.0}$	&	20 \\ 
    Ib(II) + Ib & 2.0$\pm^{1.2}_{0.9}$	& 2.2$\pm{0.9}$	&	25 \\
    Ic-6 + Ic-7 & 2.2$\pm^{3.1}_{0.9}$	& 3.2$\pm{2.4}$	&	18 \\
    Ic-3 + Ic-4 & 3.0$\pm^{0.8}_{0.7}$	& 3.0$\pm{0.7}$	&	12 \\
    XRF-SNe	&	2.8$\pm{0.6}$ & 2.8$\pm{0.6}$ & 2 \\
    GRB-SNe	&	4.3$\pm^{2.5}_{1.2}$ & 4.7$\pm{1.5}$ & 3 \\
    All & 2.4$\pm^{1.5}_{1.0}$	& 2.8$\pm{1.5}$ & 80 \\
    \hline
	 & \citet{Drout2011}   & & \\
    Ib &  -	& 2.0$\pm^{1.1}_{0.8}$ & 8 \\
    Ic & -	& 1.7$\pm^{1.4}_{0.9}$ & 11\\
    Ic-BL &	 -  & $4.7\pm^{2.3}_{1.8}$ & 4 \\
    ``Engine-driven SNe'' & -  & $3.6\pm^{2.0}_{1.6}$ & 3 \\ 
    \hline
    & \citet{Taddia2015}   & & \\
    Ib & - & 3.6$\pm{0.6}$ & 6 \\
    Ic & - & 5.75$\pm{2.09}$ & 3 \\
    Ic-BL & - & 5.39$\pm{1.30}$ & 4 \\
    \hline
     & \citet{Lyman2016}   & & \\
    IIb &-	& 2.2$\pm{0.8}$ & 9 \\
    Ib &-	& 2.6$\pm{1.1}$ & 13 \\
    Ic &-	& 3.0$\pm{2.8}$ & 8 \\
    Ic-BL &-	& 2.9$\pm{2.2}$ & 9 \\
    \hline
    & \citet{Taddia2018}   & & \\
    IIb & - & 4.3$\pm{2.0}$ & 10 \\
    Ib & - & 3.8$\pm{2.1}$ & 10 \\
    Ic & - & 2.1$\pm{1.0}$ & 11 \\
    \hline
	\multicolumn{4}{l}{$^{\dag}$See Section~\ref{sec:intro} for how the classification schemes relate}\\
    \end{tabular}
	\label{tab:massstats}
\end{table}

\section{Discussion}
\subsubsection{Average \mej\ and lack of bimodality}
The overlap in the \mej\ distributions for different SN sub-types suggests that the progenitors of many of these SNe may be similar.
If one assumes a $1.4$ \msun\ neutron star (NS) remnant\footnote{Assuming SNe do not occur when the core collapses directly to a black hole.} then \mej\ $+ M_\mathrm{NS} = M_\mathrm{COcore} < 5$ \msun\ for the vast majority of SE-SNe.
Given that there are two possible progenitor pathways, one from low mass stars stripped through binary interaction and one from high mass single stars, it may be expected that there would be two \mej\ distributions present.

The distribution of all SNe and for the SNe Ic appears unimodal. However, it was previously mentioned that \mej\ of GRB-SNe could be underestimated with this method and that spectral modelling has shown these SNe to have \mej\ $\sim 10$ \msun\ \citep{Mazzali2003,Mazzali2013,Ashall2017}. In this case a bimodality would appear with a second peak around $10$ \msun, and it may point to GRB-SNe resulting from a different population to other SE-SNe. However, as these occur in low-metallicity environments \citep{Modjaz2008,Japelj2016,Vergani2017,Japelj2018} their progenitors may still have had significant binary interaction in order to lose their H and He envelopes.

The lack of bimodality present in the distributions could result from preference for one evolutionary pathway over the other. 
That the mean ejecta mass is low and the number of SE-SNe so high compared to the rate of WR stars \citep{Smith2011} suggests that stripping of stars $<30$ \msun\ through binary interaction is the dominant pathway. 
This can be partly reconciled if one considers variations in the IMF that allow the formation of more high mass stars \citep[e.g.,][]{Schneider2018}.

\subsubsection{Linking ejecta mass to initial mass}
The ejecta masses for He-rich SNe, and $L_\mathrm{norm}$ curves from nebular spectra, are consistent with the final masses of stars with \mzams\ of $10 - 18$ \msun\ stripped in binary environments \citep{Yoon2017,Limongi2017}. 
This may also be similarly true for SNe Ic, because the difficulty in the single star scenario is in explaining the low mean ejecta mass of SNe Ic, with the condition that the mass of He in the pre-explosion envelope \mhe\ $<0.14$ \msun\ \citep{Hachinger2012}, and with the relative observed rates of SNe Ic. 

The models of \cite{Woosley1993} showed it was possible to obtain a pre-SN mass star of $\sim4$ \msun\ from a 60 \msun\ star and for it to be below the $M_\mathrm{He}$ limit.
\cite{Chieffi2013} find progenitors compatible with our criteria from rotating single stars with \mzams\ $=20 - 25$ \msun. These are He-free and have \mej \la\ 5 \msun\ assuming a NS remnant. Stars with \mzams\ greater than this have compatible \mej, towards the tail of the \mej\ distribution.

\cite{Yoon2017b} adapted the mass-loss rates of Wolf-Rayet stars and better matched the observed properties (low ejecta masses) of SNe Ic, obtaining pre-SNe stars of $\sim 5$ \msun\ from He stars of initial mass $M= 10 - 12$ \msun\ that could explode as SNe Ic with the ejecta masses we find here. \cite{Yoon2017b} state that their 15 \msun\ He stars arise from stars with \mzams $=30-40$ \msun, thus the 10--12 \msun\ He stars would be expected to have \mzams\ below this.

None of the models from \cite{Georgy2009} fulfilled both the \mej\ and \mhe\ criteria, regardless of metallicity. At explosion they were too massive and too rich in elements heavier than oxygen.
The \mzams\ $=32$ -- 40 \msun\ models of \cite{Georgy2012} that include rotation fare better, as these are within the upper bounds of the He-poor \mej\ distributions (\mej \ga\ 7 \msun) and fulfil the \mhe\ criteria. 
\cite{McClelland2016} find viable SN~Ic progenitors from a stripped 30 \msun\ star, again at the high end of our \mej\ distribution, with \mej\ \ga\ 5.5 \msun\ for different metallicities.
\cite{Groh2013b} find models that fit the \mhe\ criteria from \mzams\ $>32$ \msun\ rotating stars but which predict pre-SN masses of $M > 10$ \msun.
Even accounting for large uncertainties in determining the ejecta mass, most of the values calculated here would have to increase by a factor of 4 to match the resulting C/O core mass ($\sim 10$ \msun) of a star with \mzams\ $= 30$ \msun.
Thus, these results suggest that SE-SNe, even SNe~Ic, predominantly occur from $<30$ \mzams\ stars stripped by binary interaction.
That is not to say they arise from the same type of stars however, the work of \cite{Modjaz2011} investigated metallicities at SN explosion sites and showed that SNe Ib and SNe Ic almost certainly arise from different populations \citep[but see][]{Kuncarayakti2018}.

\subsubsection{Broader implications}
These SNe and their progenitors can have several significant impacts:

\begin{itemize}
	\item{The overlap in SNe II and SN Ib/c progenitor masses supports the view that more massive stars may not explode \citep[though there are windows of explodability;][]{Oconnor2011,Sukhbold2014,Sukhbold2018} and instead collapse directly into black holes. If so, this would result in the loss of some fraction of their elemental yields.}
    \item{Low mass binary progenitors resulting in SNe Ib/c can lead to the formation of NS binaries. The binary system is required to be close in order to strip the more massive star, leading to a larger gravitational potential, which in turn reduces the probability of the system becoming unbound when the stars explode. Consequently, SE-SNe systems may contribute to the sources of short GRBs and kilonovae \citep[see][in relation to ultra-stripped SNe]{Tauris2013,Tauris2015,Moriya2017}.}
	\item{The results of the previous section suggest that high metallicity is not required to strip the star of its outer envelope (via line-driven winds) because interaction with a companion can do so. Consequently, these events could occur earlier in the history of the Universe. }
	\item{The progenitors of SE-SNe are in the theoretical mass regime of SNe II, meaning that these stars would explode regardless. That they tend to have higher \ek\ than other SNe means that the dynamical impact of CC-SNe on the local environment in the early Universe may be underestimated.}
	\item{The pre-explosion star will be highly stripped and hot, either a WR star or a He star. These stars emit copious amounts of UV radiation \citep[e.g.,][]{Gotberg2018} which can ionize the surrounding ISM. }
	\item{For an observed fractional rate of events, the O yield per SE-SN is lower for the binary pathway (low mass progenitors) than for single stars (high mass progenitors). If the majority of SE-SNe progenitors are low mass stars, the overall O yield per CC-SN decreases.}
\end{itemize}

\section{Conclusions} \label{sec:conclusion}
This work has presented observations and analysis of 18 SE-SNe discovered between 2013 and 2018. 
These data have been mostly collected on behalf of (e)PESSTO, a Liverpool Telescope observing campaign focused on SE-SNe, and the LCO Supernova Key Project.
The sample consists of 5 H/He-rich SNe, 6 H-poor/He-rich SNe, 3 SNe Ic-6/7 and 4 SNe Ic-3/4.

Most of the events observed in this work could be described as typical; their properties fall within the broader distributions of their subtype.
There are, however, some key extremes.
The type IIb(I) SN 2013bb has a broad light curve, and its ejecta mass of $\sim 4.8$ \msun\ makes it the most massive H/He-rich SNe yet found.
Another SN of this type, SN 2017hyh, has the highest velocity lines of any H-rich SE-SNe discovered so far with $v_\mathrm{H\alpha} \sim 17000$ \kms\ at maximum light. This SN also has an extremely rapid late decay rate at \dlate\ $=0.029$ \md.
SN Ib 2016jdw has a broad light curve and also line velocities that are amongst the highest for SNe of this type. With \mej\ $\sim 4.3$ \msun, it is significantly outside the typical SN Ib ejecta mass distribution and is only matched by the SN 1999ex and peculiar Ib SN 2007uy.
The Ic-4 SN 2018ie has a short rise time (\tp $\sim 8$ d), high velocity lines ($>20000$ \kms) and is the least luminous SNe of this type known.

The \mej\ mass distribution of H-rich SNe is bimodal with two peaks around 1.9 and 3.9 \msun. The SNe associated with these peaks are typically those without/with a shock cooling tail respectively. This is associated with compact/extended progenitor envelopes.
The SN~Ic-6/7 distribution has a low mass ($\sim 2.2$ \msun) peak but a high mass tail, suggesting a large range of pre-explosion stellar masses.
A full ejecta mass distribution of 80 SE-SNe appears unimodal with median \mej\ $=2.4\pm^{1.5}_{1.0}$ \msun\ and a mean ejecta mass of $2.8\pm{1.5}$ \msun. Adding a 1.4 \msun\ NS remnant to \mej\ gives a total pre-explosion core mass of $<5$ \msun\ for the vast majority of SE-SNe. 

Eight SNe had nebular phase spectra, using \mni, $L_\mathrm{[OI]}$, and the methods of \cite{Jerkstrand2015} it was found that all these objects, with the exception of SN 2016jdw, were compatible with model progenitors of \mzams $<17$ \msun. The progenitor of SN 2016jdw was unlikely to have been much more massive than this.
Comparison of \mej\ with stellar evolution models leads to the conclusion that the vast majority of SE-SN progenitor stars have \mzams\ $<30$ \msun, and are stripped by binary interaction.
In this case, the mass and metallicity dependence on envelope stripping is reduced and suggests that these kinds of events could have occurred in the early universe. Stripped stars are copious emitters of UV radiation and will ionize their local surroundings. 

These results further the tension between progenitor masses derived from light curve fitting and spectral modelling, single star stellar evolution models, and examination of explosion environments. 
The theoretical (\mzams\ = 12 -- 25 \msun) and observational (\mzams\ = 7 -- 18 \msun) range of SN II progenitors overlaps with that found in this and previous studies, yet it is clear that the explosion environments are different (SNe Ib/c tend to be found in or closer to \HII\ regions). As has been noted, the latter suggest massive progenitors $>30$ \msun\ for many SE-SNe which appear to have low ejecta mass. This discrepancy remains unsolved and should be a focus of future investigation.
A way to reconcile or identify problems, moving away from simple 1D analytical light curve models, would be to use spectral modelling in order to investigate the abundance tomography of the SN ejecta (e.g., the elemental abundance as a function of mass and velocity) then link this back to theoretical models of pre-explosion massive stars. Fortunately, we are now entering a time where the increase in number of well sampled SE-SN means the required parameter space can be mapped.

\section*{Acknowledgements}
SJP acknowledges support from an STFC grant and is funded by H2020 ERC grant no.~758638.
CA acknowledges the support provided by the National Science Foundation under Grant No. AST-1613472.
KM is supported by the UK STFC through an Ernest Rutherford Fellowship and by H2020 ERC grant no.~758638.
The Liverpool Telescope is operated on the island of La Palma by Liverpool John Moores University in the Spanish Observatorio del Roque de los Muchachos of the Instituto de Astrofisica de Canarias with financial support from the UK Science and Technology Facilities Council. 
Based on observations made with the Nordic Optical Telescope, operated by the Nordic Optical Telescope Scientific Association at the Observatorio del Roque de los Muchachos, La Palma, Spain, of the Instituto de Astrofisica de Canarias.
Support for G.P., F.O.E and O.R is provided by the Ministry of Economy, Development, and Tourism's Millennium Science Initiative through grant IC120009, awarded to The Millennium Institute of Astrophysics, MAS.
F.O.E.\ acknowledges support from the FONDECYT grant nr.\ 11170953.
DAH, and GH are funded by NSF grant AST-1313484. This work makes use of observations from the LCO network.
N.E.R. acknowledges support from the Spanish MICINN grant ESP2017-82674-R and FEDER funds
MF is supported by a Royal Society - Science Foundation Ireland University Research Fellowship
MS acknowledges support from EU/FP7-ERC grant [615929]
This work is based (in part) on observations collected at the European Organisation for Astronomical Research in the Southern Hemisphere, Chile as part of PESSTO, (the Public ESO Spectroscopic Survey for Transient Objects Survey) ESO program 188.D-3003, 191.D-0935, 197.D-1075.
Part of the funding for GROND (both hardware as well as personnel) was generously granted from the Leibniz-Prize to Prof. G. Hasinger (DFG grant HA 1850/28-1).
A.G.-Y. is supported by the EU via ERC grant No. 725161, the Quantum Universe I-Core program, the ISF, the BSF Transformative program and by a Kimmel award.
%
%The Acknowledgements section is not numbered. Here you can thank helpful
%colleagues, acknowledge funding agencies, telescopes and facilities used etc.

%%%%%%%%%%%%%%%%%%%%%%%%%%%%%%%%%%%%%%%%%%%%%%%%%%



\bibliographystyle{mnras}
\bibliography{allbib} % if your bibtex file is called example.bib

%\begin{thebibliography}{99}
%\end{thebibliography}

%%%%%%%%%%%%%%%%%%%%%%%%%%%%%%%%%%%%%%%%%%%%%%%%%%

%%%%%%%%%%%%%%%%% APPENDICES %%%%%%%%%%%%%%%%%%%%%
%
\appendix
%

%%%%%%%%%%%%%%%%%%%%%%%%%%%%%%%%%%%%%%%%%%%%%%%%%%%%%%

% Host galaxies

%%%%%%%%%%%%%%%%%%%%%%%%%%%%%%%%%%%%%%%%%%%%%%%%%%%%%%
\section{The host galaxies}\label{sec:hosts}
This section briefly details the host galaxies of some of the SNe, where such analysis could be conducted. 
\begin{figure*}
	\centering
    \includegraphics[scale=0.3]{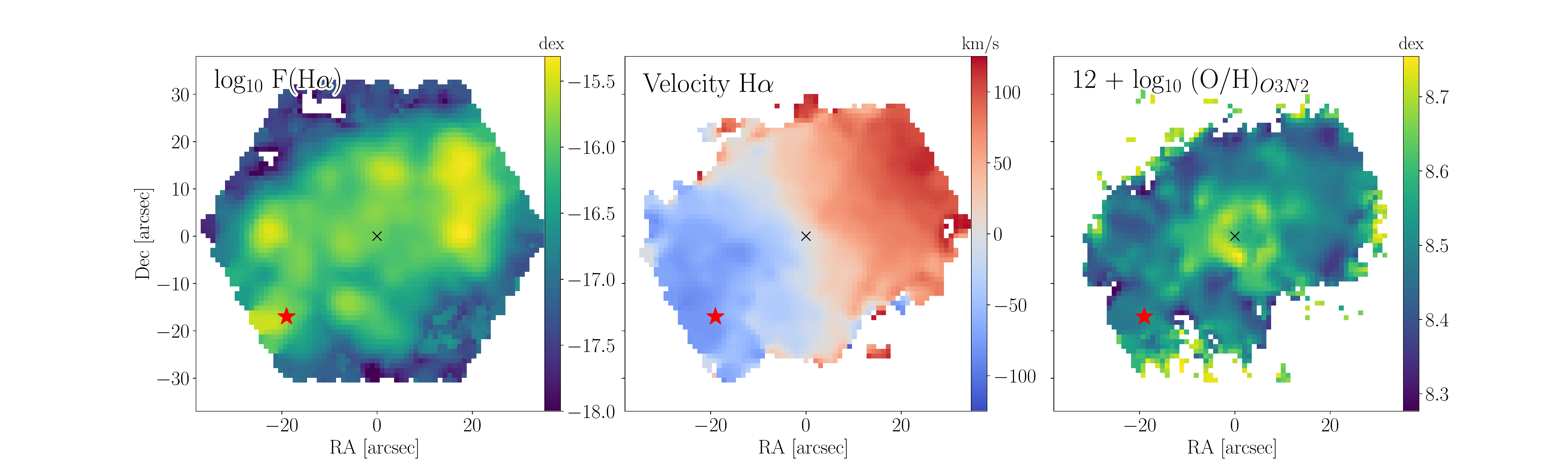}
	\caption{2MASXJ07470232+2646347, the host of SN 2017ixz in \Ha\ flux (left), \Ha\ velocity (centre), and oxygen abundance (right).}
	\label{fig:17ixzhost}
\end{figure*}

\subsubsection{SN 2013F}
The host galaxy of SN 2103F is IC\,5325, a face-on spiral of type SABbc with a heliocentric recession velocity of 1503~km\,s$^{-1}$, giving an adopted distance of 20.9~Mpc and a distance modulus of 31.60\,mag.  The galaxy has a B-band absolute magnitude M$_{B,0} = -19.8$ , which, when combined with its classification, means that it closely resembles the Milky Way.  Given that SN2013F occurred in the inner disk, just 1.9~kpc from the nucleus, the local metallicity was probably slightly higher than solar, although we are not aware of any direct measurements for this galaxy.  GALEX imaging indicates strong star formation throughout the central disk of IC\,5325.

\subsubsection{SN 2013bb}
The host galaxy of SN 2013bb, NGC 5504,  is a bright face-on spiral of type SAB(s)bc, with a extinction-corrected $B$-band absolute magnitude of --20.91\,mag.
It has strong spiral structure, and some indications of tidal disturbance, probably due to two close but fainter companions, IC~4383 and UGC~9086, which have identical recession velocities and are only 1.8 and 2.2 arcminutes in projected separation from NGC~5504, respectively.  SN 2013bb exploded 10.5~kpc in projected distance from the nucleus of NGC~5504, on the outer edge of a strong spiral arm, and hence in the outer parts of a bright galaxy with a strong similarity to the Milky Way.

\subsubsection{Type Ib SN 2013ek}

SN 2013ek occurred in NGC\,6984, a bright SBc  spiral with a recession velocity of 4670~km\,s$^{-1}$, giving a corrected Hubble flow distance of 67.5~Mpc and a distance modulus of 34.15\,mag.  The galaxy has a B-band absolute magnitude M$_{B,0} = -21.3$\,mag, and the SN occurred at a projected galactocentric radius of 3.1~kpc, leading to an inferred metallicity that is likely to be somewhat supersolar.
SN 2013ek exploded less than 0.4 arcsec from the type Ic-7 SN 2012im \citep{Mili2014AAS}.

\subsubsection{The type Ib SN 2015ah} \label{sec:15ah}

The host galaxy of SN 2015ah is a face-on late-type barred spiral, classified in NED as SABcd, with strong spiral structure, although the supernova occurred in a region between the main arms on a short spiral arm spur.  The offset from the nucleus of the galaxy corresponds to a projected radial distance of 4.0~kpc, in the central disk regions, which in terms of likely metallicity of the host environment may offset the low luminosity of the galaxy, $M_{B,0} = $ --19.29\,mag for our adopted distance.  So, even though there are no direct measurements of star formation activity or metallicity at the SN location, it appears likely to have moderate recent star formation, and approximately solar metallicity.

\subsubsection{Type Ib SN 2015ap}

The host galaxy of SN 2015ap is IC\,1776, an SBd late-type barred spiral galaxy with a heliocentric recession velocity of 3410~km\,s$^{-1}$ ($z=0.0114$), giving an inferred distance of 45.1~Mpc and a distance modulus of 33.27 mag.  The absolute B-band magnitude of the galaxy is M$_{B,0} = -19.6$\,mag, and SN2015ap occurred quite far out in the disk, 7.2~kpc from the nucleus, where the metallicity is likely to be approximately solar.  The most striking aspect of the local environment of SN 2015ap is that it lies in an extended complex of extremely vigorous star formation, which is easily the brightest such complex within IC\,1776.

\subsubsection{SN 2016P}\label{sec:16P}
NGC~5374 is a bright, strongly-barred Milky-Way like spiral, classified in NED as SBbc.  The $u$-band image of the galaxy highlights a strong star-forming ring, approximately coincident with the end of the bar; the location of SN 2016P lies directly on a bright knot within this region, giving strong evidence of current star formation which is confirmed by the narrow emission lines evident in the SN spectra.  
Further evidence for strong star formation in this ring is provided by the fact that it has hosted the two previously mentioned core-collapse SNe; the type II SN 2003bl and the He-poor SN Ic-6 2010do. The apparent magnitude of the host galaxy is $M_{B,0} =$ --21.05, and the projected radial distance of the SN from the nucleus is 6.9~kpc.  Thus this should be a region of high metallicity, similar to or slightly higher than solar.

\subsubsection{Ic-7 SN 2016iae} \label{sec:16iae}

NGC~1532 is a bright edge-on spiral galaxy, classified in NED as SBb pec. It has a strong bulge component, and the disk is clearly disturbed, probably through interaction with a very nearby, high-surface brightness amorphous companion, NGC~1531, which has a recession velocity only 65~km~S$^{-1}$ higher than that of NGC~1532. 
The location of SN 2016iae is quite central; outside the bright bulge region but on the inner disk, with a projected separation from the nucleus of 3.0~kpc.  However, the true separation is uncertain given the edge-on orientation of the galaxy. NGC~1532 is a luminous spiral galaxy, with an absolute magnitude of $M_{B,0} = $ --21.18\,mag for our adopted distance, and thus the metallicity at the SN location is likely to be high, probably somewhat above solar.
Three supernovae have been observed in NGC 1532. The Type II SN 1981A, SN 2016iae, and the Type II SN 2016ija \citep{Tartaglia2018}, which exploded serendipitously during our observing campaign.

\subsubsection{Type IIb(I) SN 2017ixz}

The host galaxy of SN 2017ixz, 2MASX J07470232+2646347, was observed on 2015 November 7th by the Mapping Nearby Galaxies at APO (MaNGA) survey \citep{Bundy2015}, Figure~\ref{fig:17ixzhost}. It was observed with the larger integral field unit (IFU) bundle, which provides with a coverage of 30$\times$30 squared arcsec, which practically covers the whole extent of the galaxy. Following \citep{Galbany2014} procedures, we analysed both the local environment of the SN within the galaxy and the integrated properties of the host. 

SN 2017ixz exploded close to a bright region in the outskirts of its host galaxy. 
We measure an absolute star formation rate of 3.681$\pm$0.122 10$^{-2}$ \msun\ yr$^{-1}$ at the SN location, an SFR density of 0.016$\pm$ 0.034 \msun\ yr$^{-1}$ kpc$^{-2}$, with a subsolar oxygen abundance of 12+log$_{10}$(O/H)=8.38$\pm$0.12 dex (in the \citealt{Marino2013} O3N2 scale) and an average stellar age of log(A) = 7.01$\pm$1.51 indicating a high content of young stars at the SN environment. 
For 2MASX J07470232+2646347 we got a total stellar mass of log(M$_\mathrm{*}$) = 10.77$\pm$0.07 dex, and the corresponding values for the oxygen abundance (in the same scale) of 8.51$\pm$0.11 dex, an SFR of 2.618$\pm$0.679 \msun\ yr$^{-1}$ and SFR density 0.004$\pm$0.001 \msun\ yr$^{-1}$ kpc$^{-2}$, and an average stellar age of log(A) = 8.32$\pm$1.26 dex. 
Our results indicate, that the SN is on a metal-poorer region compared to the average of its host (as expected from its large offset from the core; 6.9 kpc), in an environment with stars that are on average younger than in other locations of the galaxy, and with a higher rate of star formation density. All this parameters are in agreement with the typical environments of SNe IIb \citep{Galbany2018}: outskirts of the galaxy with no high SFR, and young populations with low metallicity.

% Don't change these lines
\bsp	% typesetting comment
\label{lastpage}
\end{document}